\documentclass[aps,
final,%
notitlepage,%
oneside,%
twocolumn,
nofootinbib,%
floatfix, pra]%
{revtex4-1}
\pdfoutput=1

\usepackage{color}
\newcommand{\rd}{{\mathrm d}}

\newcommand{\e}{{\mathrm e}}
\newcommand{\Ccal}{{\mathcal C}}

\newcommand{\rhoG}{\rho_{\textrm G}}

\newcommand{\vertiii}[1]{{\left\vert\kern-0.25ex\left\vert\kern-0.25ex\left\vert #1 
		\right\vert\kern-0.25ex\right\vert\kern-0.25ex\right\vert}}

\newcommand{\eq}[1]{\begin{equation} #1 \end{equation}}
\newcommand{\tr}{\mathrm{Tr}}

\newcommand{\eqarray}[1]{\begin{eqnarray} #1 \end{eqnarray}}
\newcommand{\ket}[1]{\vert #1 \rangle}
\newcommand{\bra} [1] {\langle #1 \vert}

\newcommand{\mean}[1]{\langle #1 \rangle}

\newcommand{\dd}{\mathrm{d}}

\usepackage{graphicx}
\usepackage{dcolumn}
\usepackage{bm}
\usepackage{color}
\usepackage{soul}
\usepackage{amsmath}
\usepackage[dvipsnames]{xcolor}
\usepackage{mathtools}
\usepackage{dsfont}
\usepackage{amssymb}
\usepackage[utf8]{inputenc}

\usepackage{hyperref}
\hypersetup{
	colorlinks=true,
	linkcolor=ForestGreen,
	filecolor=TealBlue,      
	urlcolor=Orchid,
	citecolor=Peach
}
\usepackage{pifont}
\usepackage{amssymb}

\usepackage{titlesec}

\begin{document}

	\title{Quadrature Coherence Scale of Linear Combinations of Gaussian Functions in Phase Space}
\author{Anaelle Hertz}
\affiliation{National Research Council of Canada, 100 Sussex Drive, Ottawa, Ontario K1N 5A2, Canada}

\author{Aaron Z. Goldberg}
\affiliation{National Research Council of Canada, 100 Sussex Drive, Ottawa, Ontario K1N 5A2, Canada}
\affiliation{Department of Physics, University of Ottawa, 25 Templeton Street, Ottawa, Ontario, K1N 6N5 Canada}

\author{Khabat Heshami}
\affiliation{National Research Council of Canada, 100 Sussex Drive, Ottawa, Ontario K1N 5A2, Canada}
\affiliation{Department of Physics, University of Ottawa, 25 Templeton Street, Ottawa, Ontario, K1N 6N5 Canada}

\affiliation{Institute for Quantum Science and Technology, Department of Physics and Astronomy,
	University of Calgary, Alberta T2N 1N4, Canada}

	\setcounter{tocdepth}{1}

	\begin{abstract}
The quadrature coherence scale (QCS) is a recently introduced measure that was shown to be an efficient witness of nonclassicality. It takes a simple form for pure and Gaussian states, but a general expression for mixed states tends to be prohibitively unwieldy. In this paper,  we introduce a method for computing the quadrature coherence scale of quantum states characterized by Wigner functions expressible as linear combinations of Gaussian functions. Notable examples within this framework include cat states, GKP states, and states resulting from Gaussian transformations, measurements, and breeding protocols. In particular, we show that the quadrature coherence scale serves as a valuable tool for examining the scalability of nonclassicality in the presence of loss. Our findings lead us to put forth a conjecture suggesting that, subject to 50\% loss or more, all pure states lose any QCS-certifiable nonclassicality. We also consider the quadrature coherence scale as a measure of quality of the output state of the breeding protocol. 
	\end{abstract}

 	\maketitle

\section{Introduction}
Counterintuitive quantum phenomena including superposition and entanglement have transformed from problems \cite{EPR1935} to curiosities \cite{Bell1964,Bell1966,KochenSpecker1967,vanFraassen1982} to features that confer advantages in multiple domains~\cite{Ekert1991,Bennettetal1993,Shor1995,Grover1996,Giovannettietal2004,Dowling2008,Aberg2014,Korzekwaetal2016}. Superposition and entanglement are based off of coherence, which underlies all interference effects. This makes the generation \cite{Breeding_Etesse_2,Breeding,GoldbergSteinberg2020}, manipulation \cite{ManiKarimipour2015,Bromleyetal2015,GarciaDiazetal2016,Buetal2017}, quantification \cite{Aberg2006arxiv,Baumgratzetal2014,Streltsovetal2015,Ranaetal2016,Yuetal2016,Rastegin2016,Hillery2016}, and measurement \cite{Girolami2014,Wangetal2017,Zhengetal2018,Xuetal2020,Wuetal2021,Yuanetal2023} of coherence essential tasks, especially when viewed through the lens of coherence as a resource \cite{Aberg2006arxiv,Baumgratzetal2014,WinterYang2016,Streltsovetal2017,Liuetal2020}.

Coherence between macroscopically distinct states is a good indicator of quantumness or nonclassicality. 
Several measures and witnesses of nonclassicality have been introduced, with their strengths and weaknesses. For example, the distance to the set of classical set \cite{Hillery1} is a precise measure of non-classicality, but is typically difficult to evaluate because it requires an optimization over a set of states. The negative volume of the Wigner function \cite{Kenfack} is also a very popular witness that can be hard to compute due to looking for small, rapid interferences via numerical integration. The latter also fails to detect many nonclassical states; e.g., squeezed states.
The quadrature coherence scale (QCS) was recently introduced as such an indicator, quantifying both the amount of coherence and the macroscopicity of the coherence in a quantum state; the QCS is a \textit{bona fide} witness of optical nonclassicality \cite{Hertz,Debievre} with the important advantage of bounding the distance to the set of classical states while being relatively easy to compute. Let us note also that for pure states QCS is equivalent to many other nonclassicality witnesses \cite{Goldbergetal23}. In addition, an interferometric scheme with two identical copies of a state suffices for measuring the QCS of the state without requiring full state tomography \cite{Griffet}, unlike the nonclassicality distance or Wigner negativity that require full tomography and then full optimization or integration; the QCS measurement was recently demonstrated on a  cloud quantum computer \cite{Goldbergetal23}. 

For states that are pure or Gaussian, the QCS is simple to compute. However, a general computation requires calculations involving multiple integrals over a state's entire Wigner function, which is essential to practical situations because non-Gaussian states are crucial to linear optical quantum computation \cite{LloydBraunstein1999,Bartlettetal2002,MariEisert2012,Walschaers21} and pure states are rare. 

In this work, we show that computation of the QCS is significantly easier for states that can be written as superpositions and convex combinations of Gaussian states, which includes non-Gaussian states that are important to quantum metrology and computation. This takes advantage of a recently developed formalism for states whose Wigner functions are linear combinations of not-necessarily-real Gaussian functions \cite{XanaduSumGaussian} and for simulating quantum optics with finite-rank superpositions of coherent states \cite{MarshallAnand2023}.

This description of states might sound restrictive, but it actually represents a large family of relevant states. As an example, Schr{\"o}dinger cat states are important non-Gaussian states used for demonstrating quantumness \cite{Schrodinger1935cat}, quantum-enhanced measurement sensitivity \cite{Zurek2001}, and error-correcting codes \cite{Cochraneetal1999} and belong to this class of superpositions of Gaussian states. They can be combined via linear optical networks and homodyne measurement can be performed on some branches while remaining in this class, allowing them to iteratively ``breed'' \cite{Breeding_Vasconcelos,TerhalWeigand2016,Breeding} into another important type of states known as Gottesman-Kitaev-Preskill (GKP) states \cite{GKPpaper}. Moreover, such states subject to the dominant noise source for photonics, viz., loss, still remain in this class, as do states subject to any Gaussian channels. In contrast, other nonclassicality measures such as quantum Fisher information that may be easy to compute for pure states must be computed from scratch with more challenging calculations when the state becomes mixed due to loss. These allow the QCS to measure the nonclassicality and thereby the quality of a state intended for fault-tolerant quantum computation as it improves via breeding and degrades via loss.

In addition, using the description of the P-function (see eq. \eqref{eq:class-state}), one can see that any state can be written as a combination of Gaussian (coherent) states, at least formally, because the coefficients in this combination need not be positive. Moreover, it was recently proven that any pure state can be represented with arbitrary precision as a \textit{finite} combination of coherent states, while all mixed states are just convex combinations of such \cite{MarshallAnand2023}. Our framework is thus fully general and allows the computation of the QCS of any state belonging to the correct class subject to any Gaussian transformations and measurements. It was developed with GKP states in mind and may prove useful throughout multimode quantum optics with continuous variables.








\color{black}

The paper is divided as follows. In the proceeding two sections, we start by introducing the main concept of quantum optics in phase space and then introduce the quadrature coherence scale. In Sec.~\ref{sec:linearCombinationGaussian}, we define the formalism related to the expression of the Wigner function of a state as a linear combination of Gaussian functions and in Sec.~\ref{sec:computingQCS} we show how to compute the QCS for such states. In Sec.~\ref{sec:examples}, we give concrete examples: we compute the QCS of cat states and GKP states, we study the scalability of the QCS through a loss channel, and we consider the QCS as a measure of fidelity for the output of a breeding protocol. In Sec. \ref{sec:othermeasures} we discuss other nonclassicality measures. We finally conclude in Sec.~\ref{sec:conclusion} and discuss some future research problems.
%

%

\section{Phase space formalism}
\label{PhaseSpaceFormalism}
In this section, we provide a brief overview of the symplectic formalism employed for continuous-variable states in quantum optics. More details can be found, for example, in \cite{weedbrook, Anaellethesis}.

A continuous-variable system is represented by $n$ modes. To each of them are associated the annihilation and creation operators $a_i$ and $a_i^\dag$ verifying the bosonic commutation relation $[a_i,a_i^\dag]=1$. We define the vector of quadratures $\boldsymbol{\hat r}=(\hat{x}_1,\hat{p}_1,\hat{x}_2,\hat{p}_2,\cdots,\hat{x}_n,\hat{p}_n)$,  where\footnote{Note that we employ units in which $\hbar=1$ throughout this paper.}
\begin{equation*}
\hat{x}_j=\frac1{\sqrt{2}}(a_j+a_j^\dag),\quad \hat{p}_j=-\frac{i}{\sqrt{2}}(a_j-a_j^\dag)\quad \forall j=1,\cdots,n.
\end{equation*}
Each quantum state $\rho$ can be described by a Wigner function
\eq{W(\boldsymbol{\alpha})=\frac1{\pi^{2n}}\int\chi(\boldsymbol z)\e^{\bar{\boldsymbol{z}}\cdot\boldsymbol{\alpha}-\boldsymbol{z}\cdot\bar{\boldsymbol{\alpha}}}d^{2n}\boldsymbol{z}},
where $\chi(\boldsymbol z)=\tr [\rho D(\boldsymbol{z})]$ is the characteristic function and $D( \boldsymbol{z})=\e^{\boldsymbol{z}\cdot a^\dag-\bar{\boldsymbol{z}} \cdot a}$ the displacement operator .
The Wigner function is normalized to $1$, but can take negative values; hence the qualification of \textit{quasi}probability distribution.

The first-order moments of a state $\rho$ constitute the displacement vector, defined as $\boldsymbol{\mu}=\mean{\boldsymbol{\hat r}}=\tr (\boldsymbol{\hat r}\rho)$, while the second moments make up the symmetric covariance matrix $\boldsymbol{\boldsymbol{\gamma}}$ whose elements are given by
\begin{equation}
\gamma_{ij}=\frac12\mean{\{\hat r_i,\hat r_j\}}-\mean{\hat r_i}\mean{\hat r_j}
\label{covmat}	
\end{equation}
where $\{\cdot,\cdot\}$ represents the anticommutator.  To be the covariance matrix of a valid quantum state, $\boldsymbol{\gamma}$ must satisfy the uncertainty principle $\boldsymbol{\gamma}+\frac{i}{2}\boldsymbol{\Omega}\geq0$,
with
$$
\boldsymbol{\Omega}=\bigoplus_{j=1}^n\begin{pmatrix}0&1\\-1&0	\end{pmatrix}.
$$
being the symplectic form.

A \textit{Gaussian} state $\rhoG$ is fully characterized by its displacement vector $\boldsymbol{\mu}$ and covariance matrix $\boldsymbol{\gamma}$. The name comes from the fact that its Wigner (and characteristic) function is a Gaussian function in the phase space:
 \eq{W_G(\boldsymbol{r})=\frac{\exp[-\frac12(\boldsymbol{r}-\boldsymbol{\mu})^T\boldsymbol{\gamma}^{-1}(\boldsymbol{r}-\boldsymbol{\mu})]}{(2\pi)^n\sqrt{\det \boldsymbol{\gamma}}} .\label{WignerGaussian}}

%

A {\itshape Gaussian transformation} is a transformation that will map a Gaussian state onto a Gaussian state. In the phase space, this translates into updating the covariance matrix and mean value of the Gaussian function as follows: 
\eq{\boldsymbol{\gamma} \rightarrow \boldsymbol{X}\boldsymbol{\gamma} \boldsymbol{X}^T+\boldsymbol{Y}, \qquad\boldsymbol{\mu}\rightarrow \boldsymbol{X}\boldsymbol{\mu}+\boldsymbol{d}\label{GaussianTransfo}.}
When the Gaussian transformation is unitary,  $X$ is a symplectic matrix and $Y=0$. This includes displacement, rotation, and squeezing operations. Another example of a Gaussian transformation is the  loss channel, where \eq{\boldsymbol{X}=\sqrt{\eta}\mathds{1}, \qquad \boldsymbol{Y}=(1-\eta)\mathds{1}/2, \qquad \boldsymbol{d}=0.\label{losschannel}}

 \section{Computation of the QCS -- General}
 
A state $\rho$ is said to be (optically) classical~\cite{Titulaer} if and only if there exists a positive function $P(\alpha)$, called the Sudarshan-Glauber P function, such that
 	\begin{equation}\label{eq:class-state}
 	\rho=\int P(\alpha) | \alpha \rangle\langle \alpha| \rd \alpha,
 	\end{equation}
 	where $\ket{\alpha}=D(\alpha)\ket{0}$ is a coherent state, $D( \alpha)$ is the displacement operator, 
 	and $|0\rangle$ is the vacuum state. Otherwise, the state is said to be nonclassical. In other words, a state is said to be nonclassical if it is not a probabilistic mixture of coherent states.
 	
It is typically challenging to determine if such a P-function exists and even more so to assess its positivity. Therefore, there is a need for defining measures and witnesses of nonclassicality. The most famous one is probably the negativity of the Wigner function: if the Wigner function of a state $\rho$ takes negative values at any point on the phase space, the state is nonclassical \cite{Kenfack}. Nevertheless, many other measures and witness exist \cite{Hillery1,Bach,Hillery3,Lee, Agarwal2,Lutkenhaus, Dodonov, Marian,Richter, Asboth, Ryl,Sperling,Killoran,Alexanian,Nair,Ryl2, Yadin, Kwon2,  Takagi18, Horoshko, Luo, Bohmann,Tan2020}.  In this work, we focus on the quadrature coherence scale (QCS), a recently introduced witness of nonclassicality \cite{Debievre,Hertz,HertzCerfDebievre,Goldbergetal23,QCS_photon_added}.

Nonclassicality is linked to the coherences present in a state, but, while many nonclassicality witnesses and measures consider the size of the coherences, the QCS also measures where those coherences are located, showcasing a different aspect of the nonclassicality of the state. As an example, the nonclassicality of a squeezed state will be detected by the QCS while it is concealed from the Wigner negativity.
	
	The QCS was calculated for several benchmark states including Fock states, squeezed thermal states, and cat states \cite{Hertz}, as well as states with more complicated QCSs like photon-added and subtracted Gaussian states~\cite{QCS_photon_added}.

 For an $n$-mode mixed state $\rho$, the QCS,  $\Ccal^2$, can be computed through the Wigner function as\footnote{Here, the vector $\mathbf{r}$ contains all the quadratures $x_j$ and $p_j$ of each mode $j$ and we  use  the correspondence $\alpha_j=\frac{x_j+ip_j}{\sqrt{2}}$. }
\eq{\Ccal^2_\rho=\frac1{2n}\frac{\int|\nabla W(\boldsymbol{r})|^2 \rm d^{2}\boldsymbol{r} }{\int|W(\boldsymbol{r})|^2 \rm d^{2}\boldsymbol{r}}.\label{QCSmixed}} 
For a pure state $\ket{\psi}$, it simplifies to the total noise (i.e. the sum of the variances) and many other indicators of nonclassicality as enumerated in Ref.~\cite{Goldbergetal23}. It is thus the trace of the covariance matrix $\boldsymbol{\gamma}$:
 \eq{\Ccal^2_{\ket\psi}=\frac1{n}\tr\boldsymbol{\gamma}=\frac1{n}\sum_{j=1}^n\left((\Delta x_j)^2+(\Delta p_j)^2\right).\label{QCSvariance}}
 
For a Gaussian state $\rho_G$, the QCS also takes a simpler form as it is proportional to the trace of the inverse of the covariance matrix:
 \eq{\Ccal^2_{\rho_G}=\frac1{4n}\tr\boldsymbol{\gamma}^{-1}.\label{QCSGaussian}}
 
 The QCS is not a measure, but a witness of nonclassicality. All classical states have $\Ccal_\rho\leq1$. Therefore, a value of the QCS greater than 1 certifies that the state is nonclassical. In addition, despite not being a proper measure of nonclassicality,  it was proven in \cite{Debievre} that the  distance\footnote{The distance is defined as D($\rho$,$\mathcal{E}_{cl})=\rm{inf}_{\sigma \in \mathcal{E}_{cl}}|||\tilde\rho-\tilde\sigma|||$ with $|||A|||=\sqrt{\mean{A|A}}$ and $\tilde\rho=\rho/\sqrt{\tr\rho^2}$.}
 D($\rho$,$\mathcal{E}_{cl}$) between the state $\rho$ and the set of nonclassical states $\mathcal{E}_{cl}$ is bounded by the QCS in the following way:
 \eq{\Ccal_\rho-1\leq D(\rho,\mathcal{E}_{cl})\leq \Ccal_\rho.\label{distance}}
 Hence, a state with $\Ccal_\rho\leq1$ will be either classical or so weakly nonclassical that it precludes certification, while $\Ccal_\rho> \Ccal_\sigma+1$ implies that a state $\rho$  is more nonclassical than $\sigma$.

    \section{Linear combination of Gaussian functions in phase space}
    \label{sec:linearCombinationGaussian}
    
    It is convenient to work only with Gaussian states, but this limits the pool of possibility. However, one can consider linear combinations of Gaussian functions that describe non-Gaussian states. This then describes a much larger family of states, yet allows us to still use many of the properties of Gaussian states.

Let $\rho$ be an $n$-mode state whose Wigner function can be written as a linear combination of Gaussian functions in phase space with mean values $\boldsymbol{\mu}_m$ and covariance matrix $\boldsymbol{\gamma}_m$:
    \eq{W(\boldsymbol{r})=\sum_m c_m G_{\boldsymbol{\mu}_m,\boldsymbol{\gamma}_m}(\boldsymbol{r})\label{WignersumGaussian},}
    with 
 \eq{G_{\boldsymbol{\mu}_m,\boldsymbol{\gamma}_m}(\boldsymbol{r})=\frac{\exp[-\frac12(\boldsymbol{r}-\boldsymbol{\mu}_m)^T\boldsymbol{\gamma}_m^{-1}(\boldsymbol{r}-\boldsymbol{\mu}_m)]}{\sqrt{\det(2 \pi\boldsymbol{\gamma}_m)}}.}
 Since a Wigner function is normalized to $1$, we have $\sum_m c_m=1$.
 Note that, if $G_{\boldsymbol{\mu}_m,\boldsymbol{\gamma}_m}$ represents a valid quantum state, then it is equivalent to Eq.~\eqref{WignerGaussian} and is real. However, $G_{\boldsymbol{\mu}_m,\boldsymbol{\gamma}_m}(x)$ can also be a complex (normalized) function, with Eq.~\eqref{WignersumGaussian} still describing a valid quantum state. More details can be found in Ref.~\cite{XanaduSumGaussian} where this definition was introduced. 
 
 Writing the Wigner function in the form of Eq.~\eqref{WignersumGaussian} is useful to describe, for example, cat states, Fock states, and GKP states. We will develop this last example in the next section. Even more interesting is that this formalism is useful in describing the resultant state when a Gaussian transformation or measurement is applied to a state of the form of Eq.~\eqref{WignersumGaussian}. Indeed, a Gaussian transformation maps a Gaussian state into another Gaussian state; hence by linearity, a state of the form Eq.~\eqref{WignersumGaussian} will still be written in this specific form after the Gaussian transformation. 
In addition, a deterministic Gaussian map (such a loss channel) will change neither the number nor the weights of the coefficients $c_m$ in the sum, and each covariance matrix $\boldsymbol{\gamma}_m$ and mean value vector $\boldsymbol{\mu}_m$ will be updated according to Eq.~\eqref{GaussianTransfo}. 
 Once again, we refer to  Ref.~\cite{XanaduSumGaussian} for more details regarding the application of Gaussian transformations to states written as a linear combination of Gaussian functions. We also draw the reader's attention to Ref.~\cite{yao2023design} for a more advanced work with linear combinations of Gaussian functions.
 
Note that, in theory, every state can be formatted in this way, but the description might involve arbitrary infinite sums or integrals that makes this formalism impractical. If one is willing to tolerate any arbitrarily small error in a description of their state, then any state can be written in this form with only a finite number of terms \cite{MarshallAnand2023}.

     \section{Computation of the QCS for a state whose Wigner function can be written as a sum of Gaussian functions}
     \label{sec:computingQCS}
    The crucial observation is that all of the integrals required for computing the QCS simplify to sums of integrals of products of Gaussians. Since products of Gaussians yield new Gaussians and integration of Gaussians is well understood, closed-form solutions for the QCS of this class of states can be established, as follows.

 The denominator of Eq.~\eqref{QCSmixed} is proportional to the purity $\tr\rho^2$ of the state, which can be computed as follows, using def. \eqref{WignersumGaussian} of the Wigner function:
 \begin{eqnarray}
\tr\rho^2&=&(2\pi)^n\int|W(\boldsymbol{r})|^2 \dd^2 \boldsymbol{r} \nonumber\\
&=&(2\pi)^n \int\left|\sum_m c_m G_{\boldsymbol{\mu}_m,\boldsymbol{\gamma}_m}(\boldsymbol{r})\right|^2 \dd^2\boldsymbol{r} \nonumber\\
&=&(2\pi)^n \int\sum_m \sum_n c_m \bar{c}_n G_{\boldsymbol{\mu}_m,\boldsymbol{\gamma}_m}(\boldsymbol{r}) G_{\bar{\boldsymbol{\mu}}_n,\bar{\boldsymbol{\gamma}}_n}(\boldsymbol{r}) \dd^2\boldsymbol{r} \nonumber\\
&=&(2\pi)^n\sum_m \sum_n c_m \bar{c}_n \nonumber\\
&&\times  \int G_{\boldsymbol{d},\boldsymbol{\Gamma}}(\boldsymbol{r}) \e^A \sqrt{\frac{\det(2\pi\boldsymbol{\Gamma})}{\det(2\pi\boldsymbol{\gamma}_m)\det(2\pi\bar{\boldsymbol{\gamma}}_n)}}\dd^2\boldsymbol{r} ,\nonumber
 \end{eqnarray}
 where we used that the product of two Gaussian functions gives a new Gaussian function (see for example Eq.~(337) in 	 Ref.~\cite{MatrixCookBook}), with
  \begin{eqnarray}
\boldsymbol{\Gamma}^{-1}&=&\boldsymbol{\gamma}_m^{-1}+\bar{\boldsymbol{\gamma}}_n^{-1}\\
\boldsymbol{d}&=&\boldsymbol{\Gamma}\left(\boldsymbol{\gamma}_m^{-1}\boldsymbol{\mu}_m+\bar{\boldsymbol{\gamma}}_n^{-1}\bar{\boldsymbol{\mu}}_n\right)\\
A&=&\frac12\left(\boldsymbol{\mu}_m^T\boldsymbol{\gamma}_m^{-1}+\bar{\boldsymbol{\mu}}_n^T\bar{\boldsymbol{\gamma}}_n^{-1}\right)\boldsymbol{\Gamma}\left(\boldsymbol{\gamma}_m^{-1}\boldsymbol{\mu}_m+\bar{\boldsymbol{\gamma}}_n^{-1}\bar{\boldsymbol{\mu}}_n\right)\qquad\nonumber\\
&&-\frac12\left(\boldsymbol{\mu}_m^T\boldsymbol{\gamma}_m^{-1}\boldsymbol{\mu}_m+\bar{\boldsymbol{\mu}}_n^T\bar{\boldsymbol{\gamma}}_n^{-1}\bar{\boldsymbol{\mu}}_n\right) .
  \end{eqnarray}
  Since the Gaussian function is normalized, $\int G_{\boldsymbol{d},\boldsymbol{\Gamma}}(\boldsymbol{r})\dd^2\boldsymbol{r}=1$. In addition, we can write $\boldsymbol{\Gamma}=\boldsymbol{\gamma}_m(\boldsymbol{\gamma}_m+\bar{\boldsymbol{\gamma}}_n)^{-1}\bar{\boldsymbol{\gamma}}_n$ \cite{MatrixCookBook}, which implies $\det\boldsymbol{\Gamma}=\det(\boldsymbol{\gamma}_m)\det(\bar{\boldsymbol{\gamma}}_n)/\det(\boldsymbol{\gamma}_m+\bar{\boldsymbol{\gamma}}_n)$.
   Thus,
   \eq{\tr\rho^2= \sum_m \sum_n  \frac{c_m \bar{c}_n \e^A}{\sqrt{\det(\boldsymbol{\gamma}_m+\bar{\boldsymbol{\gamma}}_n)}} \label{eq:purity}.}
  
  In a similar way, noting that the gradient of a Gaussian is given by $\nabla G_{\boldsymbol{\mu}_m,\boldsymbol{\gamma}_m}(\boldsymbol{r})=\boldsymbol{\gamma}_m^{-1} \left(\boldsymbol{r}-\boldsymbol{\mu}_m\right)G_{\boldsymbol{\mu}_m,\boldsymbol{\gamma}_m}(\boldsymbol{r})$,the numerator of Eq.~\eqref{QCSmixed} can be computed as follows:
   \begin{eqnarray}
\lefteqn{ \int|\nabla  W(\boldsymbol{r})|^2 \dd^2\boldsymbol{r} }\\
  &=&\int\left|\sum_m c_m \nabla G_{\boldsymbol{\mu}_m,\boldsymbol{\gamma}_m}(\boldsymbol{r})\right|^2 \dd^2\boldsymbol{r} \nonumber\\
   &=&\int\left|\sum_m c_m \boldsymbol{\gamma}_m^{-1} \left(\boldsymbol{r}-\boldsymbol{\mu}_m\right)G_{\boldsymbol{\mu}_m,\boldsymbol{\gamma}_m}(\boldsymbol{r})\right|^2 \dd^2\boldsymbol{r} \nonumber\\
  &=&\int\sum_{m,n} c_m \bar{c}_n G_{\boldsymbol{\mu}_m,\boldsymbol{\gamma}_m}(\boldsymbol{r}) G_{\bar{\boldsymbol{\mu}}_n,\bar{\boldsymbol{\gamma}}_n}(\boldsymbol{r})\nonumber\\
  &&\times \left(\boldsymbol{r}-\bar{\boldsymbol{\mu}}_n\right)^T\bar{\boldsymbol{\gamma}}_n^{-1}\boldsymbol{\gamma}_m^{-1}\left(\boldsymbol{r}-\boldsymbol{\mu}_m\right)\dd^2\boldsymbol{r} \nonumber\\
  &=& \sum_{m,n} c_m \bar{c}_n \int G_{d,\boldsymbol{\Gamma}}(\boldsymbol{r}) \e^A \sqrt{\frac{\det(2\pi\boldsymbol{\Gamma})}{\det(2\pi\boldsymbol{\gamma}_m)\det(2\pi\bar{\boldsymbol{\gamma}}_n)}}\nonumber\\
  &&\times
  \left(\boldsymbol{r}-\bar{\boldsymbol{\mu}}_n\right)^T\bar{\boldsymbol{\gamma}}_n^{-1}\boldsymbol{\gamma}_m^{-1}\left(\boldsymbol{r}-\boldsymbol{\mu}_m\right)\dd^2\boldsymbol{r} \nonumber\\
    &=& \sum_{m,n}  \frac{c_m \bar{c}_n  \e^A}{2\pi\sqrt{\det(\boldsymbol{\gamma}_m+\bar{\boldsymbol{\gamma}}_n)}}
    E[\left(\boldsymbol{r}-\bar{\boldsymbol{\mu}}_n\right)^T\bar{\boldsymbol{\gamma}}_n^{-1}\boldsymbol{\gamma}_m^{-1}\left(\boldsymbol{r}-\boldsymbol{\mu}_m\right)] .\nonumber
  \end{eqnarray}
  Here, $E[x] $ is the mean value of $x$ with respect to the probability distribution $G_{d,\boldsymbol{\gamma}}$. With Eq.~(357)  in 	 Ref.~\cite{MatrixCookBook} and a bit of algebra, we have
    \eqarray{\lefteqn{E[\left(\boldsymbol{r}-\bar{\boldsymbol{\mu}}_n\right)^T\bar{\boldsymbol{\gamma}}_n^{-1}\boldsymbol{\gamma}_m^{-1}\left(\boldsymbol{r}-\boldsymbol{\mu}_m\right)] }\\
  	&=
  \left(d-\bar{\boldsymbol{\mu}}_n\right)^T\bar{\boldsymbol{\gamma}}_n^{-1}\boldsymbol{\gamma}_m^{-1}\left(d-\boldsymbol{\mu}_m\right)+\tr[( \boldsymbol{\gamma}_m+\bar{\boldsymbol{\gamma}}_n)^{-1}].\nonumber}
  Hence,
  \eqarray{\int|\nabla W(\boldsymbol{r})|^2 \dd^2\boldsymbol{r}& =&\sum_{m,n} \frac{c_m \bar{c}_n  \e^A}{2\pi\sqrt{\det(\boldsymbol{\gamma}_m+\bar{\boldsymbol{\gamma}}_n)}} \nonumber\\
  &&\times \Big(\tr[( \boldsymbol{\gamma}_m+\bar{\boldsymbol{\gamma}}_n)^{-1}]\nonumber\\
	&&+ \left.\left(d-\bar{\boldsymbol{\mu}}_n\right)^T\bar{\boldsymbol{\gamma}}_n^{-1}\boldsymbol{\gamma}_m^{-1}\left(d-\boldsymbol{\mu}_m\right)\right).\nonumber\\
   		 \label{eq:numerator}}
  
  With Eqs.~\eqref{eq:purity} and \eqref{eq:numerator}, we can now compute the QCS of Eq.~\eqref{QCSmixed} without the need for any integration.

   \section{Examples}
   \label{sec:examples}
  \subsection{Cat states}
  Let us start with the simple example of a cat state defined as $\ket{cat}=\sqrt{\mathcal{N}}(\ket{\alpha}+\ket{-\alpha})$ where $\ket{\alpha}$ is a coherent state and $\mathcal{N}=(2+2 \e^{-2|\alpha|^2})^{-1}$ a normalization constant. The larger the value of $\alpha$, the more macroscopically distinguishable the two terms in the coherent superposition become and thus the larger we expect the QCS to be. Since it is a pure state, the QCS of a cat state can be computed as the sum of its variances (see Eq.~\eqref{QCSvariance}) and we find \cite{Hertz}
    \eq{\Ccal_{cat}^2=1+2\alpha^2\tanh|\alpha|^2.}
    
    To use our method, we write
  the cat state as the sum of four Gaussian functions  with the following parameters and coefficients \cite{XanaduSumGaussian}:
\begin{align}
&c_1=c_2=\mathcal{N},\quad c_3=c_4=\e^{-2|\alpha|^2}\mathcal{N},\nonumber\\
&\boldsymbol{\gamma}_1=\boldsymbol{\gamma}_2=\boldsymbol{\gamma}_3=\boldsymbol{\gamma}_4=\frac{1}{2}\mathds{1},\nonumber\\
&\boldsymbol{\mu}_1=-\boldsymbol{\mu}_2=\sqrt{2}(\mathcal{R}(\alpha),\mathcal{I}(\alpha)),\nonumber\\
&\boldsymbol{\mu}_3=\boldsymbol{\mu}_4^*=\sqrt{2}(i \mathcal{I}(\alpha),-i \mathcal{R}(\alpha)). \label{CoefCatState}
\end{align}  
  As a test, one can use the equations introduced in the previous section and compute
  \eqarray{
\boldsymbol{\gamma}&=&\frac14\mathds{1},\qquad
\boldsymbol{d}=\frac12(\boldsymbol{\mu}_m+\bar{\boldsymbol{\mu}}_n),\\
A&=&\frac12(\boldsymbol{\mu}_m+\bar{\boldsymbol{\mu}}_n)^T(\boldsymbol{\mu}_m+\bar{\boldsymbol{\mu}}_n)-\boldsymbol{\mu}_m^T\boldsymbol{\mu}_m-\bar{\boldsymbol{\mu}}_n^T\bar{\boldsymbol{\mu}}_n,\nonumber
  }
  which inserted into Eqs.~\eqref{eq:purity}, \eqref{eq:numerator}, and \eqref{QCSmixed} confirms that the same value of QCS is obtained. This method will be useful later when we study the effect of the loss channel on the QCS of a cat state.

 \subsection{GKP states }
 \label{sec:GKPstate}
 Let us now consider a more interesting example, that is the grid state introduced by Gottesman-Kitaev-Preskill (GKP) \cite{GKPpaper}. An ideal GKP state is represented by an infinite number of Dirac delta functions equally spaced by $2\sqrt{\pi}$ in the phase space: 
 \eqarray{\ket{0}_{GKP}&\propto&\sum_{s=-\infty}^{\infty}\ket{\sqrt{\pi\hbar}(2s)}_q\nonumber\\
 \ket{1}_{GKP}&\propto&\sum_{s=-\infty}^{\infty}\ket{\sqrt{\pi\hbar}(2s+1)}_q}
where $\ket{\cdot}_q$ denotes an eigenstate of the position quadrature.
Due to this gridlike property, GKP states can be used to correct displacement errors, making them essential resources for continuous-variable quantum computation \cite{Grimsmo2021,Tzitrin.PRXQuantum.2.040353,lachancequirion2023autonomous}

The ideal GKP state has infinite energy and cannot be normalizable. One thus needs to define finite-energy GKP states in order to deal with them. One option is to apply a Fock damping operator $E(\epsilon)=\e^{-\epsilon \hat n}$ with $\epsilon>0$ \cite{XanaduSumGaussian,XanaduComputationWithGKPstates}.
  It can then be shown that the Wigner function of the GKP states can be written as a linear combination of Gaussian functions in phase space, as defined in Eq.~\eqref{WignersumGaussian}. Let us introduce
\eq{\ket{\psi}_{GKP}=a_0\ket{0}_{GKP}+a_1\ket{1}_{GKP}.}
 Then  
 the density matrix of the GKP state can be expressed as a sum of Gaussian functions $G_{\boldsymbol{\mu}_m,\boldsymbol{\gamma}_{m}}(\boldsymbol{r})$ with covariance matrix $\boldsymbol{\gamma}_m$, mean values $\boldsymbol{\mu}_{m}$ and coefficients $c_m$ given by
\begin{align}
&\boldsymbol{\gamma}_m=\frac12\begin{pmatrix}\tanh(\epsilon)&0\\0&-\coth(\epsilon) \end{pmatrix},\label{CoefGKPState}\\
&\boldsymbol{\mu}_{m}= \frac{\sqrt{\pi }}{2} \begin{pmatrix}
 \text{sech}(\epsilon ) (2 k+2 l+s+t)\\  i \,\text{csch}(\epsilon ) (-2 k+2 l+s-t), 	\end{pmatrix},\ \text{and}\nonumber\\
 &c_m=\frac{a_s a_t^*}{\mathcal{N}}  \e^{\frac{\pi }{2}  \text{csch} (2 \epsilon ) (2 k+2 l+s+t)^2-\frac{\pi }{2}  \coth \epsilon  \left((2 k+t)^2+(2 l+s)^2\right)},\nonumber
 	\end{align}
 where $\mathcal{N}$ is such that $\sum c_m=1$,  $m=(k,l,s,t) | s,t\in\{0,1\}$ and $k,l\in\mathds{Z}$.

\begin{figure}[h!]
 	\centering
\includegraphics[width=7cm]{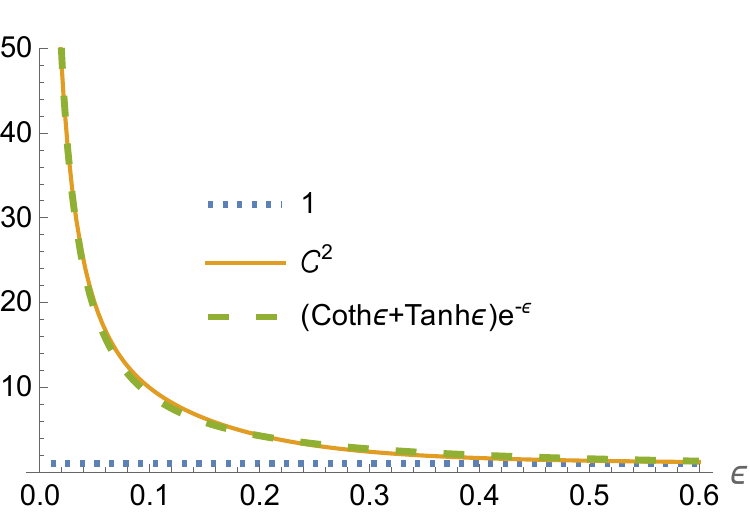}
\caption{Plot of the QCS (squared) of a GKP state as a function of the damping operator $\epsilon$. The green dashed curve shows the general behaviour of the QCS. All units here and in all subsequent figures are dimensionless.	\label{QCS_GKP_state_eq_44}}
 \end{figure}

    Using Eqs.~\eqref{eq:purity} and \eqref{eq:numerator}, we can compute the QCS of the GKP state defined above. As an example, we focus here on the state $\ket{0}_{GKP}$ ($s=t=0$, $a_0=1, a_1=0$). The result is shown in Fig.~\ref{QCS_GKP_state_eq_44} where we plotted the QCS as a function of the damping parameter $\epsilon$.
  As expected, for an ideal GKP state, that is when $\epsilon\rightarrow0$, the QCS $\rightarrow\infty$. Indeed, because they are superpositions of many different positions, GKP states have a lot of coherences located infinitely far away from each other.  The damping operator gradually erases the most distant superpositions which in turn reduces the QCS.
When $\epsilon$ is large enough, the QCS tends to 1, which means we lose all nonclassicality.
 
 The analytical formulation of the QCS is complicated, but as shown by the green curve in Fig.~\ref{QCS_GKP_state_eq_44} we can see that it evolves like 
\eq{ \Ccal^2_{GKP}\approx\e^{ -\epsilon } (\tanh (\epsilon )+\coth (\epsilon )).}

     \subsection{Evolution of the QCS through a loss channel}
     It is essential to consider loss in any actual experiment. In the domain of state generation, if the goal is to generate a nonclassical state, it is important to understand how the nonclassicality, e.g., the QCS,  evolves with losses. A simple model of loss is the loss channel whose action in the phase space is described by Eq.~\eqref{losschannel}. As mentioned in Sec.~\ref{sec:linearCombinationGaussian}, since a loss channel is described by a Gaussian transformation, when the Wigner function of the input state can be written as a sum of Gaussian functions, so will the Wigner function of the output state. The coefficient $c_m$ in Eqs.~\eqref{WignersumGaussian} remains the same and the covariance matrices and mean values are updated as follows:
     \eq{\boldsymbol{\gamma}_m(\eta)=\eta\, \boldsymbol{\gamma}_m+(1-\eta)\mathds{1}/2,\qquad \boldsymbol{\mu}_m(\eta)=\sqrt{\eta}\boldsymbol{\mu}_m.\label{upadatedParamLossChannel}}
     Hence, Eq.~ \eqref{eq:purity} and \eqref{eq:numerator} can be used to compute the QCS of the output state in terms of the loss parameter $\eta$, which allows us to analyze how the nonclassicality scales with losses.

     Loss is the dominant source of errors in photon-based protocols and can arise from multiple physical mechanisms including imperfect coupling and detector inefficiencies. It naturally arose in the reported measurements of the QCS on a cloud quantum computer~\cite{Goldbergetal23} and similarly arises in many real-life experiments with quantum states~\cite{PhysRevA.76.032309,PhysRevA.82.021801,Sperlingetal2015,Ferrettietal2024}. Fortunately, loss is additive, such that the effects of disparate loss processes can be aggregated into a single parameter $\eta$. This provides a microscopic description that, as expected, reduces the average energy in a beam by a multiplicative factor of $\eta$. Without loss, one might expect fault-tolerant quantum computation with continuous variables to already be prevalent.
  
  Let us study in more detail the evolution of the QCS through a loss channel for two specific input states: the cat states and the GKP states.

     \subsubsection{Cat states in a loss channel}

Using Eqs.~\eqref{upadatedParamLossChannel}, \eqref{QCSmixed}, \eqref{eq:purity}, and \eqref{eq:numerator}, and a little bit of algebra, it is straightforward to compute the QCS of the output of a loss channel when the input state is a cat state as defined in Eq.~\eqref{CoefCatState}.
 We obtain:
  \eq{\Ccal^2_{cat}(\eta)=1+\frac{4 \eta  | \alpha | ^2 \sinh \left(4 \alpha ^2 \eta -2 | \alpha | ^2\right)}{\cosh \left(4 \alpha ^2 \eta -2 | \alpha | ^2\right)+\cosh \left(2 | \alpha | ^2\right)+2}.}
 
  \begin{figure}[h!]
  	\centering
  	\includegraphics[width=7cm]{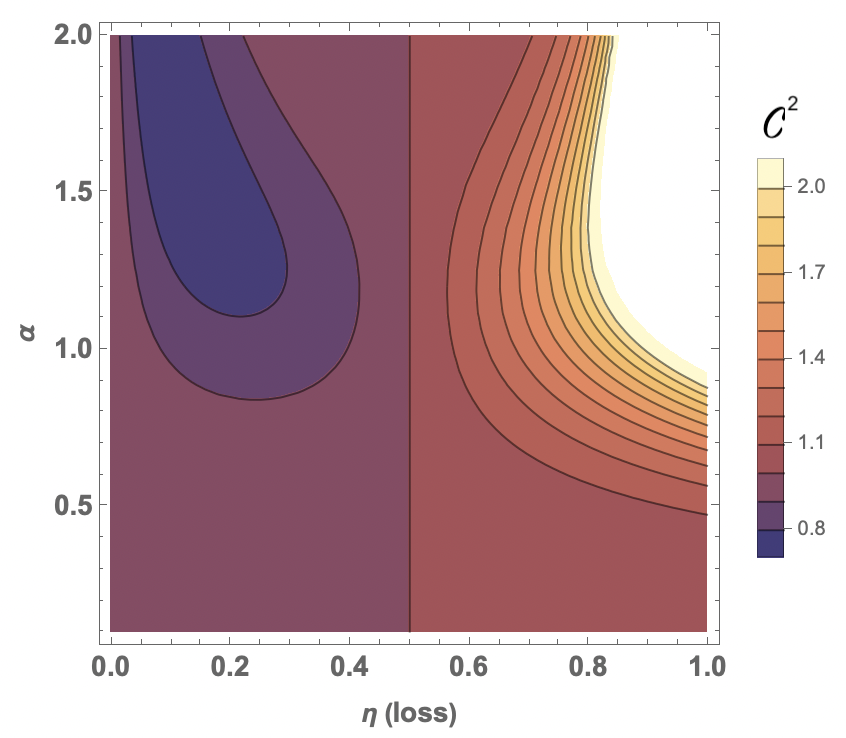}
     	\includegraphics[width=8cm]{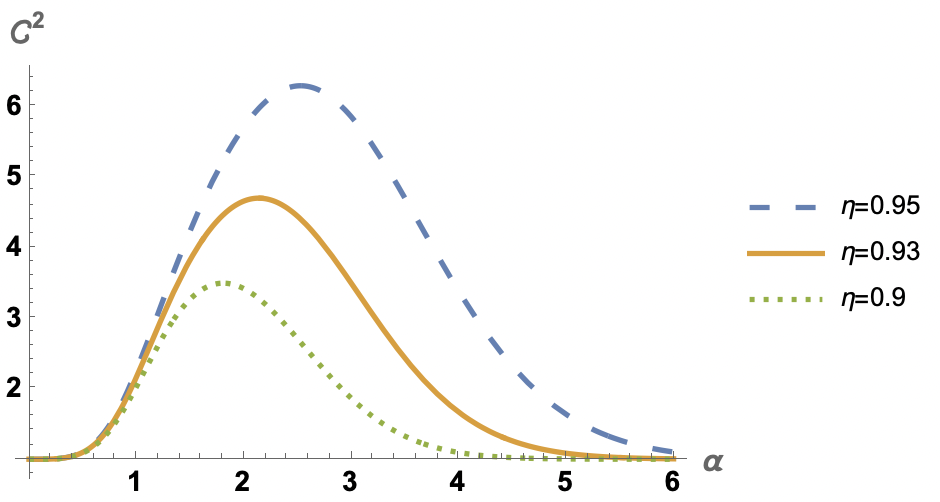}
  	\caption{Plot of the QCS (squared) of a cat state going through a loss channel, in function of $\alpha$ and $\eta$. No loss is represented by $\eta=1$. 
   \label{QCS_cat_state_loss_channel}}
  \end{figure}

  The result is plotted in Fig.~\ref{QCS_cat_state_loss_channel}. Near $\eta=1$, we can approximate the rate of losing nonclassicality as the derivative of the QCS (assuming $\alpha\in\mathds{R}$):
  \eq{\left.\frac{\partial \Ccal_{cat}^2(\eta)}{\partial_\eta}\right|_{\eta=1}=2 \alpha ^2 \left(2 \alpha ^2+\tanh \left(\alpha ^2\right)\right)=4\alpha^4-1+\Ccal^2_{cat}.}
  As already shown in \cite{Hertz}, we observe that the decoherence rate grows quadratically with the QCS (and quadratically with the energy) of the cat state: the larger the nonclassicality, the quicker we lose it.
  
  It is interesting to note that, no matter the value of $\alpha$, the QCS of the states reaches 1 (which is the nonclassicality threshold) for $\eta=1/2$.
  This phenomenon was already observed in Ref.~\cite{Goldbergetal23}.

We also remark that, in the regime of small losses ($\eta$ close to 1), the largest nonclassicality, as measured by the QCS, it not necessarily obtained with the largest value of $\alpha$. An example is given at the bottom of Fig.~\ref{QCS_cat_state_loss_channel}.  When there is no loss ($\eta=1$), the QCS tends to infinity when $\alpha$ tends to infinity; however, as soon as $\eta<1$, the QCS will tend to 1 when $\alpha\rightarrow\infty$ (and the smaller the value of $\eta$, the smallest $\alpha$ needs to be in order to reach 1). This means that, if one knows the loss of a setup available in a lab, and the aim is to create a state as nonclassical as possible (for some computing task), then $\alpha$ needs to be chosen wisely. 
Nevertheless, let us keep in mind that the QCS is not a proper measure of nonclassicality, but rather provides a bound on 
the distance D($\rho$,$\mathcal{E}_{cl}$) between the state $\rho$ and the set of classical states $\mathcal{E}_{cl}$ (see Eq.~\eqref{distance}). 
  Hence, to ensure that a state is more nonclassical in the sense that it is more distant from the set of classical states, we need a variation of QCS greater than 1. 
   When $\eta\leq1/2$, we observe that $\Ccal_{cat}^2(\eta)\leq1$, which implies that the state is always classical or weakly nonclassical. Intriguingly, per Fig.~\ref{QCS_cat_state_loss_channel},
   a cat state with $\alpha=3$ subject to 5\% loss is much further away from the set of classical states than an initially nonclassical state with $\alpha\gg 3$ subject to the same 5\% loss. More emphatically, a less quantum state with $\alpha=2$ subject to more loss $\eta=0.93$ is \textit{further} from any classical state than a more quantum state with $\alpha\gg 3$ subject to more loss $\eta=0.95$.
  


   \subsubsection{GKP states in a loss channel}
\begin{figure}[h!]
 	\centering
 	\includegraphics[width=7cm]{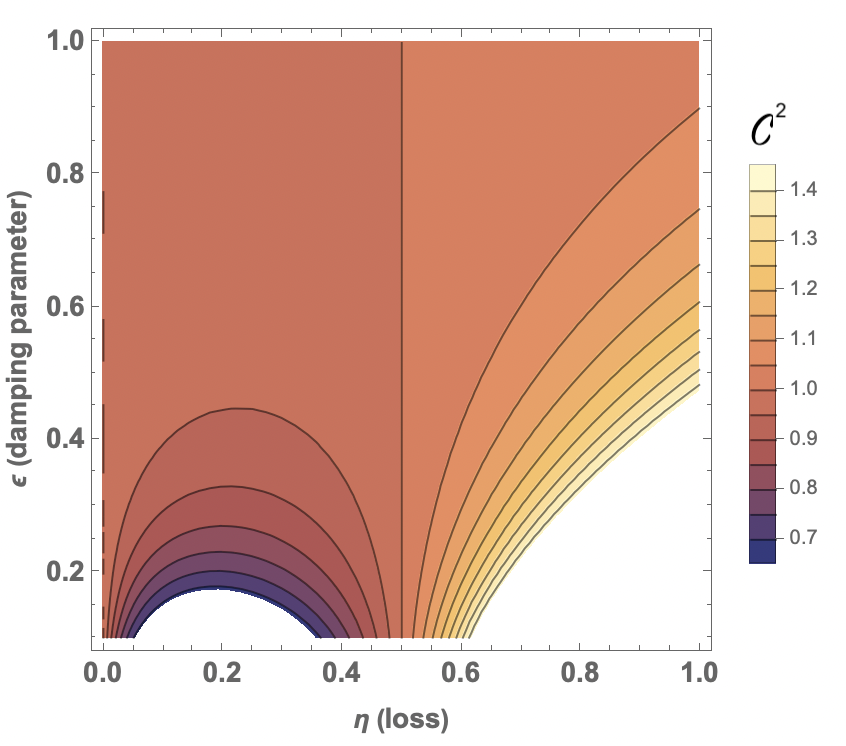}
 	 	\includegraphics[width=7cm]{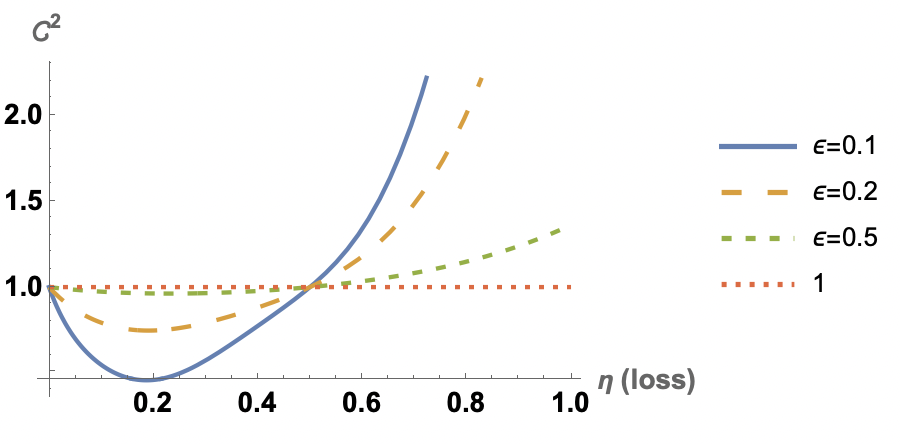}
 	\caption{Plot of the QCS (squared) of the GKP state as a function of the damping parameter $\epsilon$ and loss parameter $\eta$. No loss is represented by $\eta=1$. An ideal GKP state has $\epsilon\rightarrow 0$. \label{QCS_GKP_state_loss_channel}}
 \end{figure}

Once again, using Eqs.~\eqref{upadatedParamLossChannel}, \eqref{QCSmixed}, \eqref{eq:purity}, and \eqref{eq:numerator}, one can compute the QCS $\Ccal^2_{GKP}(\eta,\epsilon)$ of the output of a loss channel when the input state is a GKP state as defined in Eq.~\eqref{CoefGKPState}. The result is plotted in Fig.~\ref{QCS_GKP_state_loss_channel}.  No loss is represented by $\eta=1$. Hence, Fig.~\ref{QCS_GKP_state_eq_44} represents the vertical slice of the top of Fig.~\ref{QCS_GKP_state_loss_channel} when $\eta=1$.
 
As for the cat states, GKP states with better approximations (i.e. smaller value of $\epsilon$) lose quantumness more quickly with loss.
The rate of losing QCS could be evaluated by the derivative of $\Ccal^2_{GKP}(\eta)$, but since the equation is more involved, so is its derivative. Nevertheless, one can see via Fig.~\ref{QCS_GKP_state_loss_channel} that here, too, the initially more nonclassical states lose their nonclassicality more rapidly. In addition, in this case as well, the QCS of the GKP state reaches $1$ when there is 50\% loss, regardless of the initial value. However, unlike cat states, no matter the value of $\eta >1/2$, a smaller $\epsilon$ implies a larger value of the QCS. When $\eta\leq1/2$, although it varies slightly, the QCS is always smaller than or equal to 1, which implies that the state is classical or weakly nonclassical. We can consider that above $50\%$ loss (i.e. $\eta\leq 1/2$), all nonclassicality is dissipated.


 \subsubsection{Squeezed state in a loss channel}
\begin{figure}[h!]
		\centering
		\includegraphics[width=7cm]{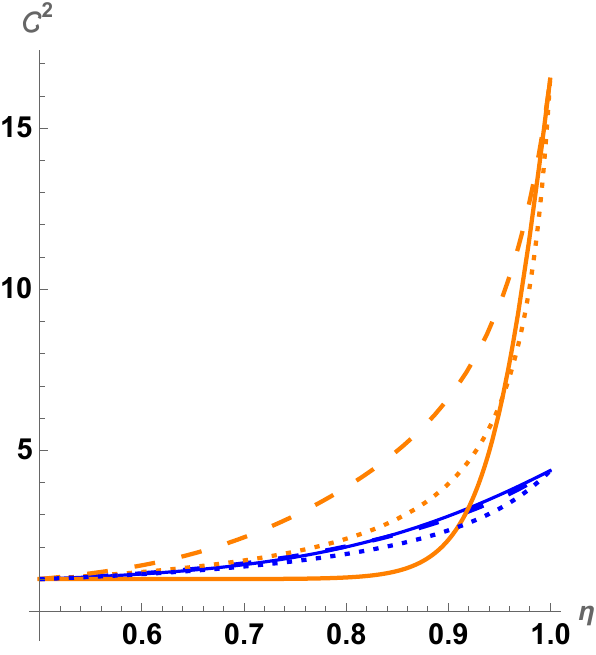}
		\caption{Comparison of the loss (parameterized by $\eta$) of nonclassicality, as measured by the QCS for a cat state (plain line), a GKP state (dashed line), and a squeezed state (dotted line). The orange curves (highest three at far right) represent states starting with $\Ccal^2\approx16.5$ (i.e. $\alpha\approx2.8$, $\epsilon\approx0.05$ and $r=1.7$)   and blue curves (lowest three at far right) states starting with $\Ccal^2\approx4.4$ (i.e. $\alpha\approx1.3$, $\epsilon\approx0.2$ and $r=1.1$). 	\label{QCS_loss_Comparison}}
	\end{figure}

	 As another example, it is also possible to compute the QCS of a lossy squeezed vacuum state $S(r)\ket{0}$, where $S(r)=\e^{\frac12(\bar r a^2-r a^{\dag2})}$ is the squeezing operator. We obtain (this result was already derived in Ref.~\cite{Goldbergetal23}):
  \eq{\Ccal^2_{Sq}(\eta)=\frac{1}{\frac{(1-2 \eta ) (\eta  \cosh (2 r)-\eta )}{-\eta +\eta  \cosh (2 r)+1}+1}.} 
	 In Fig.~\ref{QCS_loss_Comparison}, starting with different values of the QCS, we compare the loss of nonclassicality, as measured by the QCS, for GKP states (dashed line), cat states (plain line) and squeezed states (dotted line). 
We first observe that, similar to GKP states, lines representing the QCS loss of different squeezed states will never cross. This means that, even though higher initial squeezing (so high initial QCS value) will lose QCS quickly, a smaller squeezing value (thus smaller initial QCS) will always lead to lower QCS for a fixed amount of loss, even if the decay is slower.
  
Comparing the different families of states, we observe that there is no particular order for which state loses QCS faster. For large initial QCS values, GKP states are the most resilient and squeezed states lose QCS most quickly, but this conclusion does not hold anymore when initial QCS values get smaller.


Again, we see that, for all those example states, the QCS reaches $1$ at exactly $50\%$ loss. This phenomenon can be proven for any pure one-mode Gaussian state or Fock state (see Appendix \ref{Appendix}). This leads us to believe that all pure states become classical (according to the QCS) after this threshold.  It was shown in Ref.~\cite{RadimFilip} that the Wigner function of the output of a loss channel is always positive as soon as there is at least 50$\%$ loss. Indeed, the s-ordered quasiprobability function of the output of the loss channel, $P_{out}(\alpha,s)$, can be written as 
 \eq{P_{out}(\alpha,s)=\frac1{\eta}P_{in}\left(\frac{\alpha}{\sqrt{\eta}},\frac{s+\eta-1}{\eta}\right),\label{s_quasi_distribution_evolution}}
 where $P_{in}(\alpha,s)$ is the s-ordered quasiprobability function of the input. The Wigner function corresponds to $s=0$, which implies that $s'=(s+\eta-1)/\eta\leq-1$ as soon as $\eta\leq\frac12$. All s-ordered quasiporbability distributions with $s\leq-1$ are positive, which certifies the positivity of the Wigner function of the output when losses are greater than $1/2$. This does not prove that the state becomes classical according to Sudarshan-Glauber condition, but gives a hint that some quantum signature is lost at this specific point.
 Note that this conjecture does not extend to mixed states. The simplest example is to take a lossy squeezed state as the initial state: after 50\% loss, the QCS of this state will be below 1. A direct corollary of our conjecture is that all states become classical according to the QCS for loss equal to or greater than 50\%.


	 
	 \subsection{Breeding}
	
	 Several schemes exist that strive to efficiently generate GKP states in optical setups \cite{Menicucci2014,Eatonetal2019,Suetal2019,Tzitrin.PRXQuantum.2.040353,XanaduComputationWithGKPstates,Eatonetal2022,Takaseetal2023,Yanagimotoetal2023,Konnoetal2023arxiv,Crescimannaetal2023arxiv,AntenehPfister2023arxiv,Takaseetal2024arxiv}.
	 The breeding protocol \cite{Breeding_Etesse,Breeding_Etesse_2,Breeding_Sychev,Breeding_Vasconcelos,Breeding} is a procedure that gradually generates (or ``breeds'') GKP states by impinging squeezed cat states $\left(1+D(\alpha)\right) S(r)|0\rangle$ on a 50:50 beam splitter and measuring the $p$-quadrature of one output mode. By post-selecting on specific values of the $p$-quadrature, one can obtain a GKP state in the other output mode of the beam splitter. This procedure can be repeated several times in order to increase the fidelity of the output state. In one recent protocol, the vast majority of $p$-measurement values suffice for generating GKP states, singling out GKP states as ``nonclassical fixed points'' of linear optics. The result is an approximation of a GKP state with $ \sqrt{2}\alpha$ spacing and $\e^r$ squeezing.
  
	  In what is called a slow breeding protocol, the output of the first round of breeding is inserted again with another squeezed cat state of displacement $\alpha/\sqrt{2}$ at the second input, and the $p$-quadrature of one of the outputs is once again measured. This continues, where in the $M$th round the input squeezed cat state is of the form $\left(1+D(\alpha/\sqrt{2^{M-1}})\right) S(r)|0\rangle$.  More details can be found in Ref.~\cite{Breeding}. Note that, at each round, one had to do a post-selection on the value of the $p$-quadrature. As in Refs.~\cite{Breeding,Breeding_Vasconcelos, Breeding_Etesse}, we choose here to post-select on $p=0$. Nevertheless, keep in mind that the variance of the probability distribution of measuring $p=0$ scales as $\e^{2r}$. 

	 		   \begin{figure}[h!]
	 	\centering
	 	\includegraphics[width=0.45\textwidth]{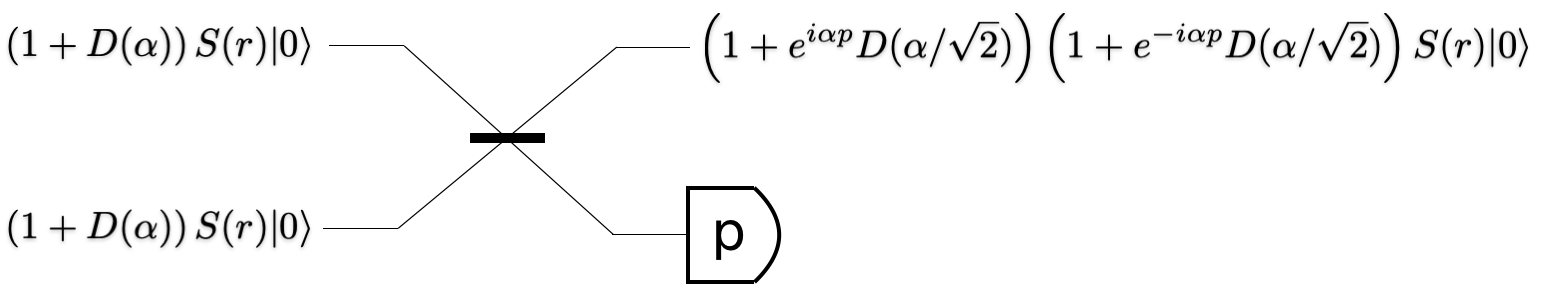}
	 	\caption{Scheme of the first round of a breeding protocol. 	\label{first_round_breeding}}
	 \end{figure} 
	 
	One important question is to understand how many rounds of breeding one needs to get a GKP state with high enough fidelity. And how can this fidelity be measured? We suggest here to look at the QCS. Indeed, we know that the QCS of a cat state grows with its amplitude $\alpha$ and we thus expect the QCS to increase through the breeding protocol. Since the output can be written as a sum of Gaussian functions and the breeding circuit is a Gaussian transformation, the QCS of the output can be easily computed with the equations presented in Sec.~\ref{sec:computingQCS}.

\begin{figure*}[t]
	 	\centering
	 	\includegraphics[width=\textwidth]{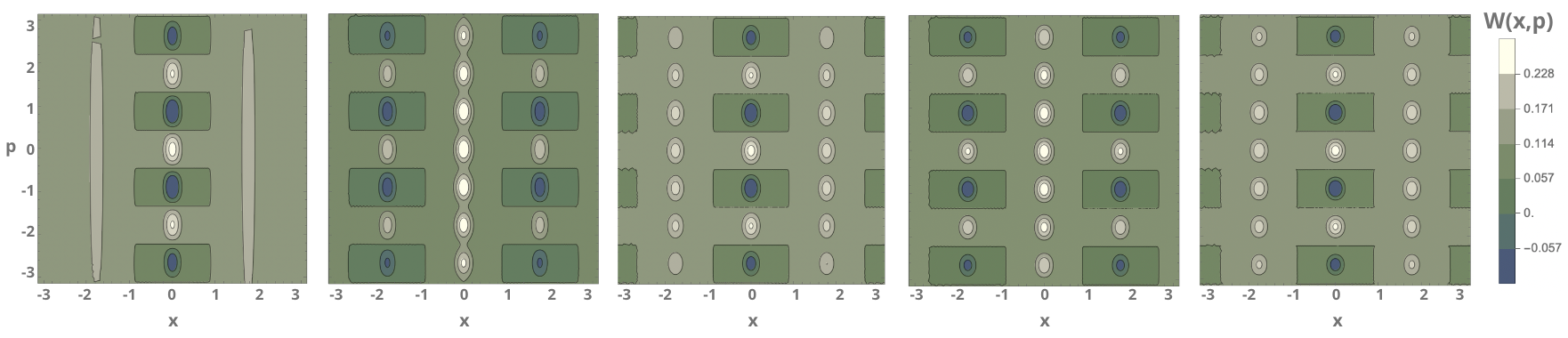}

	 	\caption{Wigner functions of a GKP state generated after M=0,1,2,3,4 rounds of the slow breeding protocol with final spacing of $2\sqrt{\pi}$ and squeezing $r=-\ln(0.2)$.	Note that at each step, we applied a final displacement of $(M+1) \sqrt{  \pi/2}$ in the $x$-direction to center the state on $0$. Colour bar indicates the magnitude and sign of the Wigner function. \label{Wigner_function_breeding}}
	 \end{figure*}

   The first round of the breeding protocol is depicted in  Fig.~\ref{first_round_breeding} and the output state is given by  $$\ket{\psi}=\left( 1+e^{i\alpha p}D\left(\frac{\alpha}{\sqrt{2}}\right)\right) \left( 1+e^{-i\alpha p}D\left(\frac{\alpha}{\sqrt{2}}\right)\right) S(r)|0\rangle.$$ To describe the output state after $M$ rounds of the slow breeding protocol, let us define the measurement operator 
	 \eq{\mathcal{M}(\phi,\alpha)=1+\e^{i\phi} D(\alpha).}	 The output state will then have the form \cite{Breeding}
	 \small
	 	 \begin{align}
&\ket{out}=\prod_{j=1}^M\mathcal{M}(\tilde\phi_j,\tilde\alpha_j)\mathcal{M}(\tilde\psi_M,\tilde\beta_M)S(r)\ket{0}\\
&=\prod_{j=1}^M\mathcal{M}\left(\phi_j+\alpha_j p,\frac{\alpha_j}{\sqrt{2}}\right)\mathcal{M}\left(\psi_M-\beta_M p,\frac{\beta_M}{\sqrt{2}}\right)S(r)\ket{0}\nonumber\\
&=\prod_{j=1}^M\mathcal{M}\left(\theta_j,\frac{\alpha}{\sqrt{2^M}}\right)\mathcal{M}\left(-\frac{\alpha p}{\sqrt{2^{M-1}}},\frac{\alpha}{\sqrt{2^M}}\right)S(r)\ket{0}\nonumber\\
&=\prod_{j=1}^M\left(1+\e^{i\theta_j p_j}D_M)\right)\left(1+\e^{-i\alpha p_M/\sqrt{2^{M-1}}}D_M)\right)S(r)\ket{0} ,\nonumber
	 	 \end{align}\normalsize
	 	 where 
	 \eqarray{	 \theta_j&=&\sum_{k=1}^M\frac{\alpha}{\sqrt{2^{k-1}}}(-1)^{(j+1)k}\nonumber\\ D_M&=&D\left(\frac{\alpha}{\sqrt{2^M}}\right).}
	 When post-selecting on $p=0$ at each step, this simplifies to 
	 \eqarray{\ket{out}&=&\sum_{k=0}^{M+1}\binom{M+1}{k}D(\alpha/\sqrt{2^M})^{k}S(r)\ket{0}\nonumber\\
	 &=&\sum_{k=0}^{M+1}\binom{M+1}{k}D(k\alpha/\sqrt{2^M})S(r)\ket{0}.\ \ }
	 The output grid state has a final spacing of $\alpha/\sqrt{2^{M-1}}$. Hence, for a GKP state with spacing $2\sqrt{\pi}$, the initial cat state must have a coherent amplitude of $\sqrt{2^{M+1}\pi}$. Fig.~\ref{Wigner_function_breeding} shows the evolution of the Wigner function after $M=0,1,2,3,4$ rounds of slow breeding. $M=0$ corresponds to a squeezed cat state.

	The density matrix of the output state has the form $\rho_{breed}=\ket{out}\bra{out}$  
 	which is indeed a sum of Gaussian functions since a  state of the form 
\eq{
D(\beta)S(r)\ket{0}\bra{0}S^\dag(r)D^\dag(\delta).}
It has the Wigner function of a normalized Gaussian of covariance matrix $\boldsymbol{\gamma}$ and mean value $\boldsymbol{\mu}$ given by (see Appendix A of Ref.~\cite{XanaduSumGaussian})
	 \eq{\boldsymbol{\gamma}=\frac12\begin{pmatrix}
\e^{-2r}&0\\0&e^{2r}
	 \end{pmatrix},\qquad\boldsymbol{\mu}=\sqrt{\frac12}\begin{pmatrix}
\beta+\delta\\i\e^{2r}(\delta-\beta)
	 \end{pmatrix},}
	 multiplied by the prefactor $\e^{-\frac12\e^{2r}(\beta-\delta)^2}$.
	 Note that, for simplicity, we assume that the displacement is real. 

  Using the equations of Sec.~\ref{sec:computingQCS}, we can compute the QCS at each round. Note that $M=0$ corresponds to the  input squeezed cat state and the QCS is given by
	 
	 \eq{\Ccal^2_{in}=\frac{1}{2} \alpha ^2 \left(1-\frac{e^{4 r}+1}{e^{\frac{1}{2} \alpha ^2 e^{2 r}}+1}\right)+\cosh (2 r).}
	
	 

	 

\begin{figure}[t!]
		\centering
		\includegraphics[width=0.85\columnwidth]{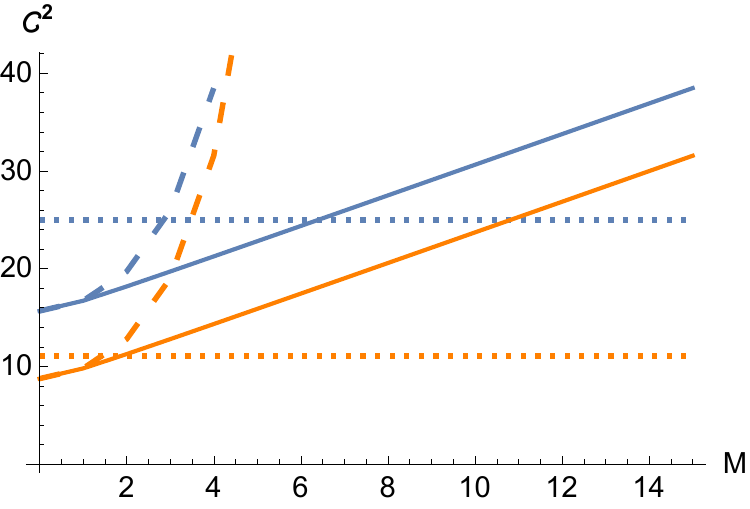}
		\caption{Evolution of the QCS (squared) after $M$ rounds of the slow (solid line) and efficient (dashed line) breeding protocol. The dotted reference line corresponds to the value of the QCS of a GKP state with spacing $2\sqrt{\pi}$ and a Fock damping operator $\tanh \epsilon=\e^{-2r}$. Here, we chose $r=-\ln(0.2)\approxeq1.6$ (blue line; upper of each pair) and $r=-\ln(0.3)\approxeq1.2$ (orange line; lower of each pair).	\label{Evolution_breeding}}
	\end{figure}

In Fig.~\ref{Evolution_breeding}, we compare the QCS of a GKP state as defined in Sec.~\ref{sec:GKPstate} with the QCS of the output state of $M$ rounds of the (slow) breeding protocol. We compare two cases: $r=-\ln(0.2)$ (blue line) as in Ref.~\cite{Breeding} and  $r=-\ln(0.3)$ (orange line). We observe that the QCS increases rapidly and reaches the "target" value (i.e. the GKP state as described in Sec.~\ref{sec:GKPstate} with $\tanh\epsilon=\e^{-2r}$) after only a few rounds. But more importantly, we see that the QCS increases with each round of breeding and thus tends to infinity, as we would expect from a perfect GKP state. This is in contrast with the GKP state as described by the damping operator, but can be explained by looking at the Wigner function.

\begin{figure}[t!]
 \centering
 \includegraphics[width=0.85\columnwidth]{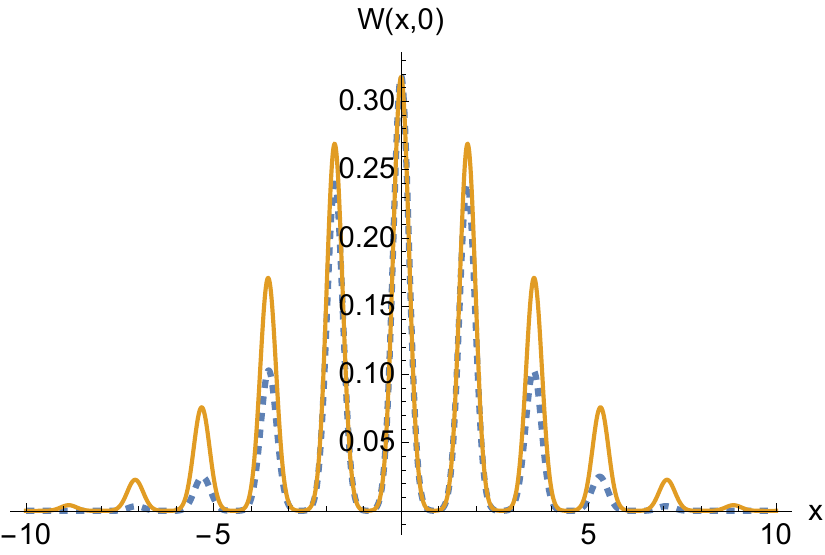}
 \caption{Comparison of the Wigner function $W(x,p=0)$ of a GKP state (dashed blue line) and the output state of 5 rounds of the breeding protocol (plain orange line). 	\label{Comparison_Wigner_GKP_Breeding}}
	\end{figure}

In Fig.~\ref{Comparison_Wigner_GKP_Breeding}, we plotted a slice of the Wigner function $W(x,p=0)$ of a GKP state ($\epsilon\approx0.09$) and the output state of 5 rounds of the breeding protocol ($r=-\ln(0.3)\approx1.2$). As we can see, the spacing of each Gaussian peak is the same, but the Gaussian envelope is not. The variance of the Gaussian envelope of the GKP state as described by the damping operator is smaller, so that the total energy of the state as well as the QCS are smaller. On the other hand, after each round of the breeding protocol, the Wigner function has more and more non-zero peaks further from the origin so that both energy and QCS tend to infinity when the number of rounds $M$ grows. This explains why the QCS of the breeding protocol does not tend to the value of the QCS of the targeted GKP state as described by the damping operator in Sec.~\ref{sec:GKPstate}. 

A more {\it efficient} scheme of breeding is realized by using a parallelized procedure. The first round is the same as the slow protocol, but then, instead of inserting a squeezed cat state in the second input of the beam splitter, we insert the same bred state to both inputs. The output state is then described by 
\eqarray{\ket{out}&=&\sum_{k=0}^{2^M}\binom{2^M}{k}D(k\alpha/\sqrt{2^M})S(r)\ket{0}.}
We now need $2^M$ input squeezed cat states for $M$ rounds of breeding (as opposed to $M+1$ inputs states for the slow protocol), but, as can be seen in Fig.~\ref{Evolution_breeding}, this scheme is much more efficient and we reach an acceptable value of QCS after only 2-3 rounds.

\section{Other measures for sums of Gaussian functions}
\label{sec:othermeasures}
Suppose we restrict our states to those whose Wigner functions can be written as sums of Gaussian functions. Does that simplify the computation of other measures of quantumness?

The general answer is negative. Consider computation of the quantum Fisher information for a state's sensitivity to displacements, averaged over all displacement directions. More quantumness means higher average quantum Fisher information, which is a metric specifically tied to the usefulness of the state for metrology. For pure states, this is simply proportional to the average number of photons in the state, the same as the QCS \cite{Goldbergetal2020extremal}. For mixed states $\rho$, such as a cat state subject to loss, the quantum Fisher information is given by~\cite{SidhuKok2020}
\begin{equation}
    \mathsf{Q}_\rho=-\int_0^\infty \dd s\tr\{([\hat{x},\rho]\e^{-\rho s})^2+([\hat{p},\rho]\e^{-\rho s})^2\}.
\end{equation} If not for the factors $\e^{-\rho s}$ and the integral over $s$, this would look very similar to the QCS (hence their agreement for pure states); that $\rho$ has a Wigner function that is a linear combination of Gaussians does not simplify this formula in any known way.

One metric that is easy to compute for states that are linear combinations of Gaussians is the total noise; i.e., the trace of the state's covariance matrix. Unfortunately, this quantity tends to be most useful for pure states, where it is equal to the QCS, but it becomes less insightful for mixed states, where expressions like the quantum Fisher information $\mathsf{Q}_\rho$ are more useful and yet do not inherit the ease of computation with linear combinations of Gaussians.

Next, consider computation of a state's negative Wigner volume
\begin{equation}
    \mathsf{N}_\rho=\int \dd^2\mathbf{r} \frac{|W(\mathbf{r})|-W(\mathbf{r})}{2}.
\end{equation} This formula is straightforward to manipulate when the state is subject to loss, as the Wigner function's transformation is known [see Eq.~\eqref{s_quasi_distribution_evolution}].
However, computing the absolute value $|W(\mathbf{r})|$ for a linear combination of Gaussian functions is challenging because each coefficient in the linear combination is, in general, complex. Moreover, the Gaussian functions themselves may be complex, and adding loss changes those complex functions, even if the complex coefficients $c_m$ are unchanged by loss.

As for distance to classical states, it is not known whether there exists a faster optimization procedure for states whose Wigner functions are linear combinations of Gaussians. It is indeed the case that the classical states belong to this category, but that does not imply that the ``straightest path'' from a given state to a given classical state is through the set of states whose Wigner functions are combinations of Gaussians. New optimizations are required for each state and each set of parameters such as loss, so this special class of states does not ameliorate the prohibitiveness of distance measures for straightforward computations.

\begin{figure}[h!]
  	\centering
      \includegraphics[width=7cm]{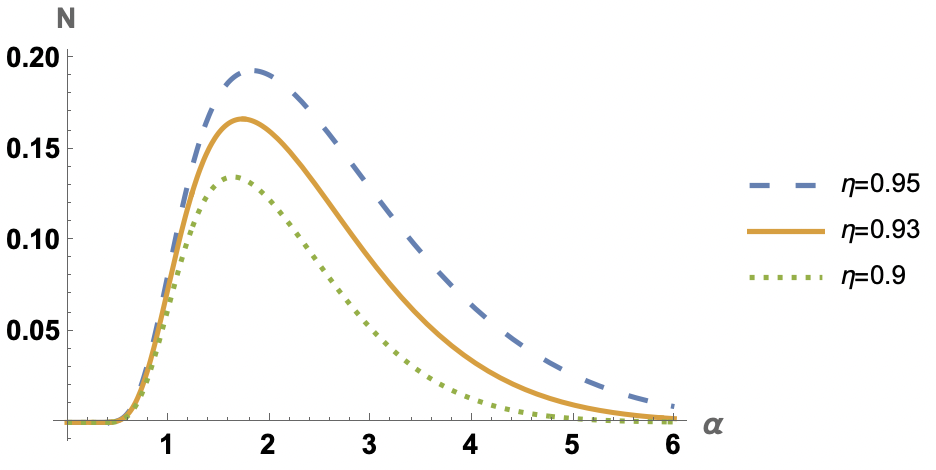}
  	\caption{Plot of $\mathsf{N}$, the negative volume of the Wigner function of a cat state going through a loss channel, as functions of $\alpha$ and $\eta$. The peak in $\mathsf{N}$ versus $\alpha$ shifts with $\eta$ in the same way as does the QCS in Fig.~\ref{QCS_cat_state_loss_channel}.\label{NegVol_Catstates}}
  \end{figure}

With that all in mind, we compare the behaviour of $\mathsf{N}_\rho$ to $\Ccal^2$ for the particular class of lossy cat states in Fig.~\ref{NegVol_Catstates}. The surprising features that we learned from the QCS again feature in the Wigner negativity, showcasing a correlation in their ability to find physical properties. Nonetheless, $\mathsf{N}$ is not known to bound the distance to classical states and thus, alone, could not be interpreted in any other way than witnessing nonclassicality.
In addition, the computation of the QCS was rapid with our formalism while the computation of $\mathsf{N}_\rho$ required a costly integration over phase space for each point on the curve, as would be required for computing the negativity of any filtered quasiprobability distribution \cite{Tan2020}. That similar information can be gleaned from each confirms the usefulness of our results.

\color{black}

\section{Conclusion}
\label{sec:conclusion}

This paper presented a general formulation of the QCS of mixed quantum states whose Wigner functions can be expressed as a linear combination of Gaussian functions.
This family of states encompasses important non-Gaussian states like Schrödinger cat states and GKP states, but also the output of a breeding protocol or, more generally, all resulting states of Gaussian operations applied to any states of such a form. Note that, in this work, we only considered transformations that are deterministic Gaussian completely positive and trace-preserving maps, but the class of Gaussian operations falling into this formalism is much larger and one could consider conditional dynamics (when the measurement of some modes updates the remaining modes), which shows the versatility and applicability of the proposed framework. With such a transformation, the number as well as the weight of each coefficient $c_m$ in Eq.~\eqref{WignersumGaussian} would change. More details can be found in Ref.~\cite{XanaduSumGaussian}.

The methods presented here could be applied to states whose Wigner functions are more complicated than linear combinations of Gaussian functions, with some modification. If some of the functions in the linear combination are given by $f(\mathbf{r})G(\mathbf{r})$ for some polynomial $f(\mathbf{r})$, there are closed-form solutions for each Gaussian integral for each monomial in the polynomials. This happens, for example, when a Gaussian state is acted on by a polynomial function of creation and annihilation operators. The expressions become more complicated as the degrees of the polynomials increase, especially when faced with linear combinations of different polynomials, but the crux of the methods here that rely on Gaussian integration can be broadly applied.

Loss as described by a loss channel falls into this framework, which allowed us to assess the scalability of nonclassicality, as measured by the QCS, with loss. As expected, for all states, the QCS decreases when loss increases, as long as the loss is less than 50\%. At this exact point, the QCS of all our examples of pure states reached a value of exactly 1, which leads us to conjecture that this is true for all pure states. We extend the conjecture by suggesting that all states lose their nonclassicality (as measured by the QCS) at the latest after 50\% loss, a statement which is supported by all Wigner functions becoming strictly non-negative at the 50\%-loss mark.

Because GKP states are so important, in particular for their error-correcting properties in quantum computing codes, we studied the breeding protocol that is one way to experimentally create GKP states and can be expressed using the formalism described in this paper. We used the QCS as a way to measure the quality of the resulting GKP state and showed that a parallelized procedure is more efficient.
In a realistic scenario, one should include loss in the protocol and understand the threshold limit of the parameters allowing to keep a sufficient amount of nonclassicality. In particular, it would be interesting to study how to optimize the squeezing parameter for state generation under realistic conditions. One could ask the following question: is it better to start with highly squeezed states (so very sensitive to loss of nonclassicality) and directly generate cat or GKP states, or is it better to input less squeezed states in a breeding protocol, which we have seen has the power to increase the nonclassicality?

\section*{Acknowledgments}

The authors acknowledge that the NRC headquarters is located on the traditional unceded territory of the Algonquin, Anishinaabe, Haudenosaunee, and Mohawk people.
 This work was supported by the NRC’s Quantum Sensors Challenge program. 
 AZG acknowledges support from an NSERC postdoctoral fellowship.
  KH acknowledges support from NSERC's Discovery Grant program.
  The authors also thank Eli Bourassa, Ilan Tzitrin and Rafael Alexander from Xanadu for fruitful discussions.

	\appendix
 \section{Appendix}	\label{Appendix}

  \subsection{QCS of a pure Gaussian state after 50\% loss}

Since the QCS is invariant under displacement, we can assume that the state is centered at the origin. A pure one-mode Gaussian state has the covariance matrix 
\eq{\boldsymbol{\gamma}=\begin{pmatrix}\sigma_{11}&\sigma_{12}\\\sigma_{12}&\sigma_{22}\end{pmatrix}}
 such that $\det(\boldsymbol{\gamma})=1/4$. After 50\% loss ($\eta=\frac12$), the covariance matrix becomes $\frac12(\boldsymbol{\gamma}+\mathds{1}/2)$. We can compute the QCS with Eq.~\eqref{QCSGaussian} :
\eqarray{\Ccal^2&=&\frac12\tr\left(\frac{\boldsymbol{\gamma}+\mathds{1}/2}{2}\right)^{-1}\nonumber\\
&=&\frac{2 (\sigma_{11} +\text{$\sigma_{22} $}+1)}{ 4 \det(\boldsymbol{\gamma}) +2\sigma_{11} +2 \text{$\sigma_{22} $}+1}=1}
and see that the result is $1$ for all one-mode pure Gaussian states.

 \subsection{QCS of a Fock state after 50\% loss}

We know from Ref.~\cite{RadimFilip} that the Wigner function of the output state of the loss channel is proportional to the Q-function of the input state (see Eq.~\eqref{s_quasi_distribution_evolution}). The Q-function of a Fock state $\ket{k}$ is given by \cite{linowski2023relating}
\eq{Q_k(x,p)=\frac1{\pi}\frac{(x^2+p^2)^k}{2^k k!}\e^{-(x^2+p^2)/2}.}
Therefore, after 50\% loss ($\eta=0.5$), the Wigner function is given by
\eq{W_k(x,p;\eta=1/2)=\frac1{\pi}\frac{(x^2+p^2)^k}{k!}\e^{-(x^2+p^2)},}
where we used the correspondence $\alpha=(x+ip)/\sqrt{2}$.
Using Eq.~\eqref{QCSmixed}, we can compute the QCS of the lossy Fock state and confirm that we obtain 1 for all $k$.

\bibliographystyle{unsrt}
	 \bibliography{references_QCS_GKP_states.bib}

\begin{thebibliography}{100}

\bibitem{EPR1935}
A.~Einstein, B.~Podolsky, and N.~Rosen.
\newblock Can quantum-mechanical description of physical reality be considered complete?
\newblock {\em Physical Review}, 47:777--780, May 1935.

\bibitem{Bell1964}
John~Stewart Bell.
\newblock On the {Einstein Podolsky Rosen} paradox.
\newblock {\em Physics}, 1(3):195--200, 1964.

\bibitem{Bell1966}
John~S. Bell.
\newblock On the problem of hidden variables in quantum mechanics.
\newblock {\em Rev. Mod. Phys.}, 38:447--452, Jul 1966.

\bibitem{KochenSpecker1967}
Simon Kochen and E.~P. Specker.
\newblock The problem of hidden variables in quantum mechanics.
\newblock {\em J. Math. and Mech.}, 17(1):59--87, 1967.

\bibitem{vanFraassen1982}
Bas~C. van Fraassen.
\newblock {The Charybdis of Realism: Epistemological Implications of {B}ell's Inequality}.
\newblock {\em Synthese}, 52(1):25--38, 1982.

\bibitem{Ekert1991}
Artur~K. Ekert.
\newblock Quantum cryptography based on {B}ell's theorem.
\newblock {\em Phys. Rev. Lett.}, 67:661--663, Aug 1991.

\bibitem{Bennettetal1993}
Charles~H. Bennett, Gilles Brassard, Claude Cr\'epeau, Richard Jozsa, Asher Peres, and William~K. Wootters.
\newblock Teleporting an unknown quantum state via dual classical and {Einstein-Podolsky-Rosen} channels.
\newblock {\em Physical Review Letters}, 70:1895--1899, Mar 1993.

\bibitem{Shor1995}
Peter~W. Shor.
\newblock Polynomial-time algorithms for prime factorization and discrete logarithms on a quantum computer.
\newblock {\em SIAM Review}, 41(2):303--332, 1999.

\bibitem{Grover1996}
Lov~K Grover.
\newblock A fast quantum mechanical algorithm for database search.
\newblock In {\em Proceedings of the twenty-eighth annual ACM symposium on Theory of computing}, pages 212--219, 1996.

\bibitem{Giovannettietal2004}
Vittorio Giovannetti, Seth Lloyd, and Lorenzo Maccone.
\newblock Quantum-enhanced measurements: Beating the standard quantum limit.
\newblock {\em Science}, 306(5700):1330--1336, 2004.

\bibitem{Dowling2008}
Jonathan~P. Dowling.
\newblock {Quantum optical metrology – the lowdown on high-N00N states}.
\newblock {\em Contemporary Physics}, 49(2):125--143, 2008.

\bibitem{Aberg2014}
Johan \AA{}berg.
\newblock Catalytic coherence.
\newblock {\em Phys. Rev. Lett.}, 113:150402, Oct 2014.

\bibitem{Korzekwaetal2016}
Kamil Korzekwa, Matteo Lostaglio, Jonathan Oppenheim, and David Jennings.
\newblock The extraction of work from quantum coherence.
\newblock {\em New Journal of Physics}, 18(2):023045, feb 2016.

\bibitem{Breeding_Etesse_2}
Jean Etesse, Martin Bouillard, Bhaskar Kanseri, and Rosa Tualle-Brouri.
\newblock Experimental generation of squeezed cat states with an operation allowing iterative growth.
\newblock {\em Phys. Rev. Lett.}, 114:193602, May 2015.

\bibitem{Breeding}
Daniel~J. Weigand and Barbara~M. Terhal.
\newblock Generating grid states from {S}chr\"odinger-cat states without postselection.
\newblock {\em Phys. Rev. A}, 97:022341, Feb 2018.

\bibitem{GoldbergSteinberg2020}
Aaron~Z. Goldberg and Aephraim~M. Steinberg.
\newblock Transcoherent states: Optical states for maximal generation of atomic coherence.
\newblock {\em Physical Review X Quantum}, 1:020306, Oct 2020.

\bibitem{ManiKarimipour2015}
Azam Mani and Vahid Karimipour.
\newblock Cohering and decohering power of quantum channels.
\newblock {\em Phys. Rev. A}, 92:032331, Sep 2015.

\bibitem{Bromleyetal2015}
Thomas~R. Bromley, Marco Cianciaruso, and Gerardo Adesso.
\newblock Frozen quantum coherence.
\newblock {\em Phys. Rev. Lett.}, 114:210401, May 2015.

\bibitem{GarciaDiazetal2016}
Mar\'{\i}a Garc\'{\i}a-D\'{\i}az, Dario Egloff, and Martin~B. Plenio.
\newblock A note on coherence power of n-dimensional unitary operators.
\newblock {\em Quantum Info. Comput.}, 16(15–16):1282–1294, nov 2016.

\bibitem{Buetal2017}
Kaifeng Bu, Asutosh Kumar, Lin Zhang, and Junde Wu.
\newblock Cohering power of quantum operations.
\newblock {\em Physics Letters A}, 381(19):1670--1676, 2017.

\bibitem{Aberg2006arxiv}
Johan {Aberg}.
\newblock {Quantifying Superposition}.
\newblock {\em arXiv e-prints}, pages quant--ph/0612146, December 2006.

\bibitem{Baumgratzetal2014}
T.~Baumgratz, M.~Cramer, and M.~B. Plenio.
\newblock Quantifying coherence.
\newblock {\em Physical Review Letters}, 113:140401, Sep 2014.

\bibitem{Streltsovetal2015}
Alexander Streltsov, Uttam Singh, Himadri~Shekhar Dhar, Manabendra~Nath Bera, and Gerardo Adesso.
\newblock Measuring quantum coherence with entanglement.
\newblock {\em Phys. Rev. Lett.}, 115:020403, Jul 2015.

\bibitem{Ranaetal2016}
Swapan Rana, Preeti Parashar, and Maciej Lewenstein.
\newblock Trace-distance measure of coherence.
\newblock {\em Phys. Rev. A}, 93:012110, Jan 2016.

\bibitem{Yuetal2016}
Xiao-Dong Yu, Da-Jian Zhang, G.~F. Xu, and D.~M. Tong.
\newblock Alternative framework for quantifying coherence.
\newblock {\em Phys. Rev. A}, 94:060302, Dec 2016.

\bibitem{Rastegin2016}
Alexey~E. Rastegin.
\newblock Quantum-coherence quantifiers based on the {T}sallis relative $\ensuremath{\alpha}$ entropies.
\newblock {\em Phys. Rev. A}, 93:032136, Mar 2016.

\bibitem{Hillery2016}
Mark Hillery.
\newblock Coherence as a resource in decision problems: The {D}eutsch-{J}ozsa algorithm and a variation.
\newblock {\em Phys. Rev. A}, 93:012111, Jan 2016.

\bibitem{Girolami2014}
Davide Girolami.
\newblock Observable measure of quantum coherence in finite dimensional systems.
\newblock {\em Phys. Rev. Lett.}, 113:170401, Oct 2014.

\bibitem{Wangetal2017}
Yi-Tao Wang, Jian-Shun Tang, Zhi-Yuan Wei, Shang Yu, Zhi-Jin Ke, Xiao-Ye Xu, Chuan-Feng Li, and Guang-Can Guo.
\newblock Directly measuring the degree of quantum coherence using interference fringes.
\newblock {\em Phys. Rev. Lett.}, 118:020403, Jan 2017.

\bibitem{Zhengetal2018}
Wenqiang Zheng, Zhihao Ma, Hengyan Wang, Shao-Ming Fei, and Xinhua Peng.
\newblock Experimental demonstration of observability and operability of robustness of coherence.
\newblock {\em Phys. Rev. Lett.}, 120:230504, Jun 2018.

\bibitem{Xuetal2020}
Huichao Xu, Feixiang Xu, Thomas Theurer, Dario Egloff, Zi-Wen Liu, Nengkun Yu, Martin~B. Plenio, and Lijian Zhang.
\newblock Experimental quantification of coherence of a tunable quantum detector.
\newblock {\em Phys. Rev. Lett.}, 125:060404, Aug 2020.

\bibitem{Wuetal2021}
Kang-Da Wu, Alexander Streltsov, Bartosz Regula, Guo-Yong Xiang, Chuan-Feng Li, and Guang-Can Guo.
\newblock Experimental progress on quantum coherence: Detection, quantification, and manipulation.
\newblock {\em Advanced Quantum Technologies}, 4(9):2100040, 2021.

\bibitem{Yuanetal2023}
Yuan Yuan, Xufeng Huang, Yueping Niu, and Shangqing Gong.
\newblock Optimal estimation of quantum coherence by bell state measurement: A case study.
\newblock {\em Entropy}, 25(10), 2023.

\bibitem{WinterYang2016}
Andreas Winter and Dong Yang.
\newblock Operational resource theory of coherence.
\newblock {\em Phys. Rev. Lett.}, 116:120404, Mar 2016.

\bibitem{Streltsovetal2017}
Alexander Streltsov, Gerardo Adesso, and Martin~B. Plenio.
\newblock Colloquium: Quantum coherence as a resource.
\newblock {\em Reviews of Modern Physics}, 89:041003, Oct 2017.

\bibitem{Liuetal2020}
Yunchao Liu and Xiao Yuan.
\newblock Operational resource theory of quantum channels.
\newblock {\em Phys. Rev. Res.}, 2:012035, Feb 2020.

\bibitem{Hillery1}
Mark Hillery.
\newblock Nonclassical distance in quantum optics.
\newblock {\em Phys. Rev. A}, 35:725--732, Jan 1987.

\bibitem{Kenfack}
Anatole Kenfack and Karol Zyczkowski.
\newblock Negativity of the {W}igner function as an indicator of non-classicality.
\newblock {\em Journal of Optics B: Quantum and Semiclassical Optics}, 6(10):396--404, aug 2004.

\bibitem{Hertz}
Anaelle Hertz and Stephan De~Bi\`evre.
\newblock Quadrature coherence scale driven fast decoherence of bosonic quantum field states.
\newblock {\em Phys. Rev. Lett.}, 124:090402, Mar 2020.

\bibitem{Debievre}
Stephan De~Bi\`evre, Dmitri~B. Horoshko, Giuseppe Patera, and Mikhail~I. Kolobov.
\newblock Measuring nonclassicality of bosonic field quantum states via operator ordering sensitivity.
\newblock {\em Phys. Rev. Lett.}, 122:080402, Feb 2019.

\bibitem{Goldbergetal23}
Aaron~Z. Goldberg, Guillaume~S. Thekkadath, and Khabat Heshami.
\newblock Measuring the quadrature coherence scale on a cloud quantum computer.
\newblock {\em Phys. Rev. A}, 107:042610, Apr 2023.

\bibitem{Griffet}
C\'elia Griffet, Matthieu Arnhem, Stephan De~Bi\`evre, and Nicolas~J. Cerf.
\newblock Interferometric measurement of the quadrature coherence scale using two replicas of a quantum optical state.
\newblock {\em Phys. Rev. A}, 108:023730, Aug 2023.

\bibitem{LloydBraunstein1999}
Seth Lloyd and Samuel~L. Braunstein.
\newblock Quantum computation over continuous variables.
\newblock {\em Phys. Rev. Lett.}, 82:1784--1787, Feb 1999.

\bibitem{Bartlettetal2002}
Stephen~D. Bartlett, Barry~C. Sanders, Samuel~L. Braunstein, and Kae Nemoto.
\newblock Efficient classical simulation of continuous variable quantum information processes.
\newblock {\em Phys. Rev. Lett.}, 88:097904, Feb 2002.

\bibitem{MariEisert2012}
A.~Mari and J.~Eisert.
\newblock Positive {W}igner functions render classical simulation of quantum computation efficient.
\newblock {\em Phys. Rev. Lett.}, 109:230503, Dec 2012.

\bibitem{Walschaers21}
Mattia Walschaers.
\newblock Non-{G}aussian quantum states and where to find them.
\newblock {\em PRX Quantum}, 2:030204, Sep 2021.

\bibitem{XanaduSumGaussian}
J.~Eli Bourassa, Nicol\'as Quesada, Ilan Tzitrin, Antal Sz\'ava, Theodor Isacsson, Josh Izaac, Krishna~Kumar Sabapathy, Guillaume Dauphinais, and Ish Dhand.
\newblock Fast simulation of bosonic qubits via {G}aussian functions in phase space.
\newblock {\em PRX Quantum}, 2:040315, Oct 2021.

\bibitem{MarshallAnand2023}
Jeffrey Marshall and Namit Anand.
\newblock Simulation of quantum optics by coherent state decomposition.
\newblock {\em Optica Quantum}, 1(2):78--93, Dec 2023.

\bibitem{Schrodinger1935cat}
E.~Schr{\"o}dinger.
\newblock Die gegenw{\"a}rtige situation in der quantenmechanik.
\newblock {\em Naturwissenschaften}, 23(48):807--812, Nov 1935.

\bibitem{Zurek2001}
Wojciech~Hubert Zurek.
\newblock {Sub-Planck structure in phase space and its relevance for quantum decoherence}.
\newblock {\em Nature}, 412(6848):712--717, 2001.

\bibitem{Cochraneetal1999}
P.~T. Cochrane, G.~J. Milburn, and W.~J. Munro.
\newblock Macroscopically distinct quantum-superposition states as a bosonic code for amplitude damping.
\newblock {\em Phys. Rev. A}, 59:2631--2634, Apr 1999.

\bibitem{Breeding_Vasconcelos}
H.~M. Vasconcelos, L.~Sanz, and S.~Glancy.
\newblock All-optical generation of states for ``encoding a qubit in an oscillator''.
\newblock {\em Opt. Lett.}, 35(19):3261--3263, Oct 2010.

\bibitem{TerhalWeigand2016}
B.~M. Terhal and D.~Weigand.
\newblock Encoding a qubit into a cavity mode in circuit qed using phase estimation.
\newblock {\em Phys. Rev. A}, 93:012315, Jan 2016.

\bibitem{GKPpaper}
Daniel {G}ottesman, Alexei {K}itaev, and John {P}reskill.
\newblock Encoding a qubit in an oscillator.
\newblock {\em Phys. Rev. A}, 64:012310, Jun 2001.

\bibitem{weedbrook}
Christian Weedbrook, Stefano Pirandola, Ra\'ul Garc\'{\i}a-Patr\'on, Nicolas~J. Cerf, Timothy~C. Ralph, Jeffrey~H. Shapiro, and Seth Lloyd.
\newblock {G}aussian quantum information.
\newblock {\em Rev. Mod. Phys.}, 84:621--669, May 2012.

\bibitem{Anaellethesis}
Anaelle Hertz.
\newblock {\em Exploring continuous-variable entropic uncertainty relations and separability criteria in quantum phase space}.
\newblock PhD thesis, 2018.

\bibitem{Titulaer}
U.~M. Titulaer and R.~J. {G}lauber.
\newblock Correlation functions for coherent fields.
\newblock {\em Phys. Rev.}, 140:B676--B682, Nov 1965.

\bibitem{Bach}
A.~Bach and U.~Luxmann-Ellinghaus.
\newblock The simplex structure of the classical states of the quantum harmonic oscillator.
\newblock {\em Commun. Math. Phys.}, 107, 1986.

\bibitem{Hillery3}
Mark Hillery.
\newblock Total noise and nonclassical states.
\newblock {\em Phys. Rev. A}, 39:2994--3002, Mar 1989.

\bibitem{Lee}
Ching~Tsung Lee.
\newblock Theorem on nonclassical states.
\newblock {\em Phys. Rev. A}, 52:3374--3376, Oct 1995.

\bibitem{Agarwal2}
G.~S. Agarwal and K.~Tara.
\newblock Nonclassical character of states exhibiting no squeezing or sub-poissonian statistics.
\newblock {\em Phys. Rev. A}, 46:485--488, Jul 1992.

\bibitem{Lutkenhaus}
N.~L\"utkenhaus and Stephen~M. Barnett.
\newblock Nonclassical effects in phase space.
\newblock {\em Phys. Rev. A}, 51:3340--3342, Apr 1995.

\bibitem{Dodonov}
V.~V. Dodonov, O.~V. Man'ko, V.~I. Man'ko, and A.~Wünsche.
\newblock Hilbert-{S}chmidt distance and non-classicality of states in quantum optics.
\newblock {\em Journal of Modern Optics}, 47(4):633--654, 2000.

\bibitem{Marian}
Paulina Marian, Tudor~A. Marian, and Horia Scutaru.
\newblock Quantifying nonclassicality of one-mode {G}aussian states of the radiation field.
\newblock {\em Phys. Rev. Lett.}, 88:153601, Mar 2002.

\bibitem{Richter}
Th. Richter and W.~Vogel.
\newblock Nonclassicality of quantum states: A hierarchy of observable conditions.
\newblock {\em Phys. Rev. Lett.}, 89:283601, Dec 2002.

\bibitem{Asboth}
J\'anos~K. Asb\'oth, John Calsamiglia, and Helmut Ritsch.
\newblock Computable measure of nonclassicality for light.
\newblock {\em Phys. Rev. Lett.}, 94:173602, May 2005.

\bibitem{Ryl}
S.~Ryl, J.~Sperling, E.~Agudelo, M.~Mraz, S.~K\"ohnke, B.~Hage, and W.~Vogel.
\newblock Unified nonclassicality criteria.
\newblock {\em Phys. Rev. A}, 92:011801, Jul 2015.

\bibitem{Sperling}
J.~Sperling and W.~Vogel.
\newblock Convex ordering and quantification of quantumness.
\newblock {\em Physica Scripta}, 90(7):074024, june 2015.

\bibitem{Killoran}
N.~Killoran, F.~E.~S. Steinhoff, and M.~B. Plenio.
\newblock Converting nonclassicality into entanglement.
\newblock {\em Phys. Rev. Lett.}, 116:080402, Feb 2016.

\bibitem{Alexanian}
Moorad Alexanian.
\newblock Non-classicality criteria: {G}lauber–{S}udarshan {P} function and {M}andel parameter.
\newblock {\em Journal of Modern Optics}, 65(1):16--22, 2018.

\bibitem{Nair}
Ranjith Nair.
\newblock Nonclassical distance in multimode bosonic systems.
\newblock {\em Phys. Rev. A}, 95:063835, Jun 2017.

\bibitem{Ryl2}
S.~Ryl, J.~Sperling, and W.~Vogel.
\newblock Quantifying nonclassicality by characteristic functions.
\newblock {\em Phys. Rev. A}, 95:053825, May 2017.

\bibitem{Yadin}
Benjamin Yadin, Felix~C. Binder, Jayne Thompson, Varun Narasimhachar, Mile Gu, and M.~S. Kim.
\newblock Operational resource theory of continuous-variable nonclassicality.
\newblock {\em Phys. Rev. X}, 8:041038, Dec 2018.

\bibitem{Kwon2}
Hyukjoon Kwon, Kok~Chuan Tan, Tyler Volkoff, and Hyunseok Jeong.
\newblock Nonclassicality as a quantifiable resource for quantum metrology.
\newblock {\em Phys. Rev. Lett.}, 122:040503, Feb 2019.

\bibitem{Takagi18}
Ryuji Takagi and Quntao Zhuang.
\newblock Convex resource theory of non-{G}aussianity.
\newblock {\em Phys. Rev. A}, 97:062337, Jun 2018.

\bibitem{Horoshko}
D.~B. Horoshko, S.~De~Bi\`evre, G.~Patera, and M.~I. Kolobov.
\newblock Thermal-difference states of light: Quantum states of heralded photons.
\newblock {\em Phys. Rev. A}, 100:053831, Nov 2019.

\bibitem{Luo}
Shunlong Luo and Yue Zhang.
\newblock Quantifying nonclassicality via {W}igner-{Y}anase skew information.
\newblock {\em Phys. Rev. A}, 100:032116, Sep 2019.

\bibitem{Bohmann}
Martin Bohmann, Elizabeth Agudelo, and Jan Sperling.
\newblock Probing nonclassicality with matrices of phase-space distributions.
\newblock {\em {Quantum}}, 4:343, October 2020.

\bibitem{Tan2020}
Kok~Chuan Tan, Seongjeon Choi, and Hyunseok Jeong.
\newblock Negativity of quasiprobability distributions as a measure of nonclassicality.
\newblock {\em Phys. Rev. Lett.}, 124:110404, Mar 2020.

\bibitem{HertzCerfDebievre}
Anaelle Hertz, Nicolas~J. Cerf, and Stephan De~Bi\`evre.
\newblock Relating the entanglement and optical nonclassicality of multimode states of a bosonic quantum field.
\newblock {\em Phys. Rev. A}, 102:032413, Sep 2020.

\bibitem{QCS_photon_added}
Anaelle Hertz and Stephan De~Bi\`evre.
\newblock Decoherence and nonclassicality of photon-added and photon-subtracted multimode {G}aussian states.
\newblock {\em Phys. Rev. A}, 107:043713, Apr 2023.

\bibitem{yao2023design}
Yuan Yao, Filippo Miatto, and Nicolás Quesada.
\newblock On the design of photonic quantum circuits, 2023.

\bibitem{MatrixCookBook}
K.~B. Petersen and M.~S. Pedersen.
\newblock The matrix cookbook, nov 2012.
\newblock Version 20121115.

\bibitem{Grimsmo2021}
Arne~L. Grimsmo and Shruti Puri.
\newblock Quantum error correction with the {G}ottesman-{K}itaev-{P}reskill code.
\newblock {\em PRX Quantum}, 2:020101, Jun 2021.

\bibitem{Tzitrin.PRXQuantum.2.040353}
Ilan Tzitrin, Takaya Matsuura, Rafael~N. Alexander, Guillaume Dauphinais, J.~Eli Bourassa, Krishna~K. Sabapathy, Nicolas~C. Menicucci, and Ish Dhand.
\newblock Fault-tolerant quantum computation with static linear optics.
\newblock {\em PRX Quantum}, 2:040353, Dec 2021.

\bibitem{lachancequirion2023autonomous}
Dany Lachance-Quirion, Marc-Antoine Lemonde, Jean~Olivier Simoneau, Lucas St-Jean, Pascal Lemieux, Sara Turcotte, Wyatt Wright, Am\'elie Lacroix, Jo\"elle Fr\'echette-Viens, Ross Shillito, Florian Hopfmueller, Maxime Tremblay, Nicholas~E. Frattini, Julien Camirand~Lemyre, and Philippe St-Jean.
\newblock Autonomous quantum error correction of {G}ottesman-{K}itaev-{P}reskill states.
\newblock {\em Phys. Rev. Lett.}, 132:150607, Apr 2024.

\bibitem{XanaduComputationWithGKPstates}
Ilan Tzitrin, J.~Eli Bourassa, Nicolas~C. Menicucci, and Krishna~Kumar Sabapathy.
\newblock Progress towards practical qubit computation using approximate {G}ottesman-{K}itaev-{P}reskill codes.
\newblock {\em Phys. Rev. A}, 101:032315, Mar 2020.

\bibitem{PhysRevA.76.032309}
Metin Sabuncu, Ladislav Mi\ifmmode~\check{s}\else \v{s}\fi{}ta, Jarom\'{\i}r Fiur\'a\ifmmode~\check{s}\else \v{s}\fi{}ek, Radim Filip, Gerd Leuchs, and Ulrik~L. Andersen.
\newblock Nonunity gain minimal-disturbance measurement.
\newblock {\em Phys. Rev. A}, 76:032309, Sep 2007.

\bibitem{PhysRevA.82.021801}
Mikael Lassen, Lars~Skovgaard Madsen, Metin Sabuncu, Radim Filip, and Ulrik~L. Andersen.
\newblock Experimental demonstration of squeezed-state quantum averaging.
\newblock {\em Phys. Rev. A}, 82:021801, Aug 2010.

\bibitem{Sperlingetal2015}
J.~Sperling, M.~Bohmann, W.~Vogel, G.~Harder, B.~Brecht, V.~Ansari, and C.~Silberhorn.
\newblock Uncovering quantum correlations with time-multiplexed click detection.
\newblock {\em Phys. Rev. Lett.}, 115:023601, Jul 2015.

\bibitem{Ferrettietal2024}
Hugo Ferretti, Y.~Batuhan Yilmaz, Kent Bonsma-Fisher, Aaron~Z. Goldberg, Noah Lupu-Gladstein, Arthur O.~T. Pang, Lee~A. Rozema, and Aephraim~M. Steinberg.
\newblock Generating a 4-photon tetrahedron state: toward simultaneous super-sensitivity to non-commuting rotations.
\newblock {\em Optica Quantum}, 2(2):91--102, Apr 2024.

\bibitem{RadimFilip}
Radim Filip.
\newblock {G}aussian quantum adaptation of non-{G}aussian states for a lossy channel.
\newblock {\em Phys. Rev. A}, 87:042308, Apr 2013.

\bibitem{Menicucci2014}
Nicolas~C. Menicucci.
\newblock Fault-tolerant measurement-based quantum computing with continuous-variable cluster states.
\newblock {\em Phys. Rev. Lett.}, 112:120504, Mar 2014.

\bibitem{Eatonetal2019}
Miller Eaton, Rajveer Nehra, and Olivier Pfister.
\newblock Non-{G}aussian and {G}ottesman–{K}itaev–{P}reskill state preparation by photon catalysis.
\newblock {\em New Journal of Physics}, 21(11):113034, nov 2019.

\bibitem{Suetal2019}
Daiqin Su, Casey~R. Myers, and Krishna~Kumar Sabapathy.
\newblock Conversion of {G}aussian states to non-{G}aussian states using photon-number-resolving detectors.
\newblock {\em Phys. Rev. A}, 100:052301, Nov 2019.

\bibitem{Eatonetal2022}
Miller Eaton, Carlos Gonz{\'{a}}lez-Arciniegas, Rafael~N. Alexander, Nicolas~C. Menicucci, and Olivier Pfister.
\newblock Measurement-based generation and preservation of cat and grid states within a continuous-variable cluster state.
\newblock {\em {Quantum}}, 6:769, July 2022.

\bibitem{Takaseetal2023}
Kan Takase, Kosuke Fukui, Akito Kawasaki, Warit Asavanant, Mamoru Endo, Jun-ichi Yoshikawa, Peter van Loock, and Akira Furusawa.
\newblock {G}ottesman-{K}itaev-{P}reskill qubit synthesizer for propagating light.
\newblock {\em npj Quantum Information}, 9(1):98, 2023.

\bibitem{Yanagimotoetal2023}
Ryotatsu Yanagimoto, Rajveer Nehra, Ryan Hamerly, Edwin Ng, Alireza Marandi, and Hideo Mabuchi.
\newblock Quantum nondemolition measurements with optical parametric amplifiers for ultrafast universal quantum information processing.
\newblock {\em PRX Quantum}, 4:010333, Mar 2023.

\bibitem{Konnoetal2023arxiv}
Shunya Konno, Warit Asavanant, Fumiya Hanamura, Hironari Nagayoshi, Kosuke Fukui, Atsushi Sakaguchi, Ryuhoh Ide, Fumihiro China, Masahiro Yabuno, Shigehito Miki, Hirotaka Terai, Kan Takase, Mamoru Endo, Petr Marek, Radim Filip, Peter van Loock, and Akira Furusawa.
\newblock Logical states for fault-tolerant quantum computation with propagating light.
\newblock {\em Science}, 383(6680):289--293, 2024.

\bibitem{Crescimannaetal2023arxiv}
Valerio Crescimanna, Aaron~Z. Goldberg, and Khabat Heshami.
\newblock Seeding {G}aussian boson samplers with single photons for enhanced state generation.
\newblock {\em Phys. Rev. A}, 109:023717, Feb 2024.

\bibitem{AntenehPfister2023arxiv}
Amanuel {Anteneh} and Olivier {Pfister}.
\newblock {Machine learning for efficient generation of universal hybrid quantum computing resources}.
\newblock {\em arXiv e-prints}, page arXiv:2310.03130, October 2023.

\bibitem{Takaseetal2024arxiv}
Kan {Takase}, Fumiya {Hanamura}, Hironari {Nagayoshi}, J.~Eli {Bourassa}, Rafael~N. {Alexander}, Akito {Kawasaki}, Warit {Asavanant}, Mamoru {Endo}, and Akira {Furusawa}.
\newblock {Generation of Flying Logical Qubits using Generalized Photon Subtraction with Adaptive {G}aussian Operations}.
\newblock {\em arXiv e-prints}, page arXiv:2401.07287, January 2024.

\bibitem{Breeding_Etesse}
Jean Etesse, Rémi Blandino, Bhaskar Kanseri, and Rosa Tualle-Brouri.
\newblock Proposal for a loophole-free violation of bell's inequalities with a set of single photons and homodyne measurements.
\newblock {\em New Journal of Physics}, 16(5):053001, may 2014.

\bibitem{Breeding_Sychev}
Demid~V. Sychev, Alexander~E. Ulanov, Anastasia~A. Pushkina, Matthew~W. Richards, Ilya~A. Fedorov, and Alexander~I. Lvovsky.
\newblock Enlargement of optical {S}chr\"{o}dinger's cat states.
\newblock {\em Nat, Potonis}, 11:379--382, May 2017.

\bibitem{Goldbergetal2020extremal}
Aaron~Z. Goldberg, Andrei~B. Klimov, Markus Grassl, Gerd Leuchs, and Luis~L. Sánchez-Soto.
\newblock {Extremal quantum states}.
\newblock {\em AVS Quantum Science}, 2(4):044701, 11 2020.

\bibitem{SidhuKok2020}
Jasminder~S. Sidhu and Pieter Kok.
\newblock {Geometric perspective on quantum parameter estimation}.
\newblock {\em AVS Quantum Science}, 2(1):014701, 02 2020.

\bibitem{linowski2023relating}
Tomasz Linowski and \L{}ukasz Rudnicki.
\newblock Relating the {G}lauber-sudarshan, {W}igner, and {H}usimi quasiprobability distributions operationally through the quantum-limited amplifier and attenuator channels.
\newblock {\em Phys. Rev. A}, 109:023715, Feb 2024.

\end{thebibliography}

\end{document}